\documentclass[reprint,aps,prd,superscriptaddress,nofootinbib,floats,showpacs]{revtex4-2}
\usepackage[utf8]{inputenc}
\usepackage{graphicx}
\usepackage{amssymb}
\usepackage{textcomp}
\usepackage{amsmath}
\usepackage{tabularx}
\usepackage{bm}
\usepackage{times}
\usepackage{color}
\usepackage{ulem}
\usepackage{slashed}
\usepackage{multirow}
\usepackage{verbatim}
\usepackage{cancel}
\usepackage{subfigure}
\usepackage{epstopdf}
\usepackage{tcolorbox}
\usepackage{mathrsfs}
\usepackage{amsmath}
\usepackage{xcolor}
\usepackage[export]{adjustbox}
\def\ba{\begin{align}}
\def\ea{\end{align}}

\usepackage{tikz}
\usetikzlibrary{shapes,snakes}
\numberwithin{equation}{section}
\usepackage[colorlinks=true, pdfstartview=FitV, linkcolor=blue, citecolor=blue, urlcolor=blue]{hyperref}
\allowdisplaybreaks[4]

\def\lsim{\mathrel{\raise.3ex\hbox{$<$\kern-.75em\lower1ex\hbox{$\sim$}}}}
\def\gsim{\mathrel{\raise.3ex\hbox{$>$\kern-.75em\lower1ex\hbox{$\sim$}}}}

\definecolor{orange}{rgb}{1,0.5,0}

\begin{document}

\title{Deparametrization and quantization of scalar-tensor gravity and its cosmological model}

\author{Faqiang Yuan}
\affiliation{
School of Physics and Astronomy, Key Laboratory of Multiscale Spin Physics (Ministry of Education), Beijing Normal University, Beijing 100875, China
}
\author{Haida Li}
\affiliation{
Department of Physics, South China University of Technology, Guangzhou 510641, China
}
\author{Shengzhi Li}
\affiliation{
School of Physics and Astronomy, Key Laboratory of Multiscale Spin Physics (Ministry of Education), Beijing Normal University, Beijing 100875, China
}
\author{Yongge Ma}
\email{mayg@bnu.edu.cn}
\affiliation{
School of Physics and Astronomy, Key Laboratory of Multiscale Spin Physics (Ministry of Education), Beijing Normal University, Beijing 100875, China
}
\begin{abstract}
The degree of freedom of the scalar field in scalar-tensor gravity is employed as "time" to deparametrize the Hamiltonian constraint of the theory. The deparametrized system is then nonperturbatively quantized by the approach of loop quantum gravity. This results in a discrete time evolution of the physical states with respect  to the gravitational degree of freedom in the quantum theory. In the corresponding Brans-Dicke cosmological model, the physical solutions to the quantum Hamiltonian constraint are obtained in the light of the deparametrization. The quantum dynamics indicates that the classical big bang singularity is replaced by a quantum bounce. 
\end{abstract}

\maketitle

%%%%%%%%%%%%%%%%%%%%%%%%%%%%%%%%
\section{Introduction}
\label{sec:Intro}
%%%%%%%%%%%%%%%%%%%%%%%%%%%%%%%%

As a nonperturbatively approach to quantize gravity, loop quantum gravity (LQG) is well known with its background independence\cite{ashtekar2004background,rovelli2004quantum,han2007fundamental,thiemann2008modern,rovelli1994physical}. The construction of LQG depends on the connection-dynamical formulation of general relativity (GR). From the theoretical viewpoint, it is desirable to study whether the methods of LQG are applicable to other gravity theories alternative to GR. From the observational viewpoint, there is evidence implying that GR might need to be modified. In particular, the astronomical observations in late twentieth century, including the cosmic microwave background radiation and the large-scale structure of the Universe, indicate that the Universe is experiencing an accelerated expansion\cite{riess1998observational,spergel2007three,blake2009wigglez,ho2008correlation}. There are two predominant perspectives regarding the accelerated expansion of the Universe.
The first keeps the fundamental framework of GR while introducing a form of matter or energy with unique properties that have yet to be observed, commonly referred to as "dark energy"\cite{frieman2008dark,copeland2006dynamics}. The second perspective posits that GR is inadequate for explaining the accelerating expansion of the Universe. Although it is well tested within the Solar System, it is not necessarily applicable to the cosmological scale\cite{will2014confrontation,will2005post,will2005einstein}. There might exist gravitational theories that are consistent with GR in the Solar System yet capable of addressing the accelerating expansion of the Universe at larger scales\cite{nojiri2007introduction}. The scalar-tensor theory of gravity was originally developed by Brans and Dicke in 1961, which was supposed to match the Mach principle better than GR by introducing a nonminimally coupled scalar field to represent a variable "gravitational constant"\cite{brans1961mach}. This theory has garnered much attention, since some of its cosmological models can predict the late time accelerated expansion of the Universe\cite{sen2001late,ildes2023analytic}. It should be noted that the scalar field in the scalar-tensor theory should not be regarded as a matter field but a part of the gravitational field. In comparison to the gravity theory involving a minimally coupled scalar field, the theory with a nonminimally coupled scalar field better aligns with the experimental observations of the cosmic microwave background radiation when applied to early Universe inflation~\cite{budhi2019inflation,hertzberg2010inflation}.

The LQG framework has been extended to the scalar-tensor gravity\cite{zhang2011nonperturbative}. Since the connection dynamics of GR is governed by three first-class constraints, this leads to the picture of frozen time evolution in both the classical and the quantum theories\cite{thiemann2008modern}. To understand this picture, the concept of relative evolution has been proposed\cite{rovelli1991observable,rovelli1991quantum,rovelli1990quantum,thiemann2006solving}. Deparametrization serves as a method to implement the concept of relative evolution. In the context of GR minimally coupled with a scalar field, the scalar field can be employed as the "time" to deparametrize the Hamiltonian constraint\cite{domagala2010gravity,lewandowski2016loop,alesci2015hamiltonian}. An important issue in the deparametrization scheme is whether one can realize the relative evolution with respect to certain gravitational degree of freedom. Therefore, it is desirable to study the possibility to utilize the scalar field itself as time to deparametrize the Hamiltonian constraint of scalar-tensor gravity. In this paper, the Hamiltonian constraint of scalar-tensor gravity will be deparametrized by the nonminimally coupled scalar field, and the theory will be nonperturbatively quantized within the framework of LQG. 

The application of LQG to the cosmological models of GR has resulted in the development of a symmetry-reduced sector known as loop quantum cosmology\cite{ashtekar2003mathematical,ashtekar2011loop,ashtekar2006quantum0,ashtekar2006quantum}. This theory provides a new picture of the evolution of the early Universe where the big bang singularity in classical theory is avoided by a quantum bounce. In this paper, we will also consider the cosmological model of the original scalar-tensor theory, the so-called Brans-Dicke theory\cite{brans1961mach}. We will apply the deparametrized Brans-Dicke theory to the spatially flat, homogeneous and isotropic cosmological model, and demonstrate the quantum dynamics of the loop quantum cosmological model. 

The structure of this paper is as follows. In Sec. II, we will review the Hamiltonian analysis of scalar-tensor theory and its connection-dynamical formalism. In Sec. III, we will deparametrize the scalar-tensor theory and perform loop quantization of the deparametrized Hamiltonian constraint. In Sec. IV, we will apply our deparametrized theory to the Brans-Dicke cosmology and construct the evolution picture. Conclusions and outlooks will be given in the last section. Throughout the paper, the Greek letters $\mu, \nu, \rho ...$  represent spacetime indices, lowercase Latin letters $a, b, c... g$ represent space indices, and we set $8 \pi G=1$ and $\hbar=1$ for convenience.

%%%%%%%%%%%%%%%%%%%%%%%%%%%%%%%%%%%%%%%%%%%%%%
\section{The structure of Scalar-Tensor theory}
\label{sst}
In this section, we first revisit the Hamiltonian analysis of scalar-tensor theory and its connection-dynamical formalism. This review aims to explain the origin of each quantity in the final connection dynamical formalism, which will be used for deparametrization and loop quantization.
\subsection{Hamiltonian analysis}
\label{sec:local}

We start from the following most general action of scalar-tensor theory\cite{fujii2003scalar}:
\begin{align}
\label{action}
S(\phi,g)=\!\int d^{4} x \sqrt{-g}\left[\frac{1}{2}\left(\phi \mathcal{R}-\frac{\omega(\phi)}{\phi}\left(\partial_{\mu} \phi\right) \partial^{\mu} \phi\right)\!-\xi(\phi)\right],
\end{align}
where $g$ is the determinant of spacetime metric $g_{\mu \nu}$, $\mathcal{R}$ is the scalar curvature, and $\omega(\phi)$ and $\xi(\phi)$ can be arbitrary functions of scalar field $\phi$. To cast the theory into canonical form, one assumes that the spacetime manifold has the topology $\mathcal{M} \cong \mathbb{R} \times \Sigma$, where $\Sigma$ is a fixed three-dimensional manifold of arbitrary topology. This topology is guaranteed by the requirement that the spacetime is globally hyperbolic. By doing 3+1 decomposition of the spacetime, the induced metric on the spatial manifold $\Sigma$ reads $h_{\mu \nu}=g_{\mu \nu}+n_{\mu}n_{\nu}$, where $n_{\mu}$ is the unit normal covector of $\Sigma$ satisfying $n^{\mu}n_{\mu}=-1$. The time evolution vector field $t^{\mu}$, satisfying $t^{\mu} \nabla_{\mu} t=1$, can be decomposed into orthogonal and tangential components to the hypersurfaces, such that $t^{\mu}=N n^{\mu}+N^{\mu}$, where $N$ is called the lapse function and $N^{\mu}$ is called the shift vector. In this formalism, the spacetime metric takes the form
\begin{align}
\label{spacetimemetric}
 \! d s^2 \!=\! g_{\mu \nu} d x^\mu d x^\nu \!=\!-\! N^2 d t^2 \!+ \! h_{a b}\left(d x^a \!+ \! N^a d t\right)\!\left(d x^b \!+ \! N^b d t\right).
\end{align}
Eq.~(\ref{spacetimemetric}) indicates $\sqrt{-g}=N \sqrt{h}$, where $h$ is determinant of $h_{a b}$. Let $D_{a}$ be the spatial covariant derivative operator compatible with the spatial metric by $D_{a} h_{bc}=0$, and let $K_{a b}=D_{a} n_{b}$ be the extrinsic curvature of hypersurface $\Sigma$. Since the extrinsic curvature satisfies $K_{a b}=\frac{1}{2} \mathcal{L}_{\vec{n}} h_{a b}$, the time derivative of the spatial metric can be obtained as
\begin{align}
\label{qdot}
 \mathcal{L}_{\vec{t}} h_{a b}=\dot{h}_{a b}=2 N K_{a b}+D_{a} N_{b}+D_{b} N_{a}.
\end{align}
Then, the four-dimensional scalar curvature can be expressed as
\begin{align}
\label{scalarcurvature}
\mathcal{R}\!=\! R \!+\! K_{a b} K^{a b}\!-\! K^{2}\!-\! 2 \nabla_{\mu}\! \left(n^{\nu} \nabla_{\nu} n^{\mu}\!-\! n^{\mu} \nabla_{\nu} n^{\nu}\! \right),
\end{align}
where $K=h^{a b}K_{a b}$ and $R$ is the three-dimensional scalar curvature of $h_{ab}$. In order to define the configuration variables and their conjugate momentum variables in the Hamiltonian formulation, one needs to express the action in terms of $(\phi,N,N^{a},h_{a b},\dot{\phi},\dot{N},\dot{N}^{a},\dot{h}_{a b})$ and their spatial derivative. A straightforward calculation yields the Lagrangian:
\begin{equation}
	\label{Lagrangian}
	\begin{aligned}
		&L = \int d^{3}x\, \biggl\{N\sqrt{h}\Big[
		\frac{1}{2}\phi\left(K_{ab}K^{ab}-K^{2}+R\right)
		\\
		&\!-\! \frac{K}{N}\left(\dot{\phi}-N^{c}\partial_{c}\phi\right)\!-\! D^{a}D_{a}\phi
		\!-\! \frac{\omega(\phi)}{2\phi}(\partial_{\mu}\phi)(\partial^{\mu}\phi)
		\!-\! \xi(\phi)
		\Big]\biggr\}.
	\end{aligned}
\end{equation}
Therefore, the associated conjugate momentum variables of the configuration variables $h_{ab}$ and $\phi$ are defined, respectively, as 
\begin{align}
\label{conjugate momentum}
p^{a b}:=\frac{\partial \mathcal{L}}{\partial \dot{h}_{a b}}=\frac{\sqrt{h}}{2}\left[\phi\left(K^{a b}\!-\! K h^{a b}\right)\!-\! \frac{h^{a b}}{N}\left(\dot{\phi}\!-\! N^{c} \partial_{c} \phi\right)\right],
\end{align}
\begin{align}
\label{conjugate momentumphi}
 \pi:=\frac{\partial \mathcal{L}}{\partial \dot{\phi}}=-\sqrt{h}\left(K-\frac{\omega(\phi)}{N \phi}\left(\dot{\phi}-N^{c} \partial_{c} \phi\right)\right).
\end{align}
Since the Lagrangian $L$ does not contain $\dot{N}$ and $\dot{N^{a}}$, we obtain two primary constraints:
\begin{align}
\label{primary constraints}
M=\frac{\partial \mathcal{L}}{\partial \dot{N}}=0,\ M^{a}=\frac{\partial \mathcal{L}}{\partial \dot{N^{a}}}=0.
 \end{align}
The combination of $p\equiv h_{ab} p^{ab}$ and $\pi$ gives
 \begin{align}
\label{source of conformal constraint}
 (3+2 \omega(\phi))\left(\dot{\phi}-N^{a} \partial_{a} \phi\right)=\frac{2 N}{\sqrt{h}}(\pi \phi -p).
  \end{align}
It is clear from the above equation that $\phi,\pi, h_{ab}$ and $p^{ab}$ are not independent when $\omega(\phi)=-\frac{3}{2}$. In this case, there is a new constraint $S=p-\pi \phi =0$. Therefore, it is natural to divide the scalar-tensor theory into two sectors: one for $\omega(\phi) \neq-\frac{3}{2}$ and another for $\omega(\phi)=-\frac{3}{2}$. 

In the sector of $\omega(\phi) \neq-\frac{3}{2}$, by the Legendre transformation $H_{t}=\int d^{3} x \left(p^{a b} \dot{h}_{a b}+\pi \dot{\phi}\right)-L$, we can express the canonical Hamiltonian as
\begin{equation}
	\begin{aligned}
		\label{canonical Hamiltonian}
		H_{t} &= \int d^{3} x\, N\frac{2}{\sqrt{h}}\left(\frac{p_{a b} p^{a b}-\frac{1}{2} p^{2}}{\phi}+\frac{(p-\phi \pi)^{2}}{2 \phi(3+2 \omega)}\right) \\
		&\quad + \int d^{3}x\, N\frac{1}{2} \sqrt{h}\biggl(-\phi R+\frac{\omega(\phi)}{\phi}\left(D_{a} \phi\right) D^{a} \phi \\
		&\qquad + 2 D_{a} D^{a} \phi+2 \xi(\phi)\biggr) \\
		&\quad + \int d^{3}x\, N^{a}\left(-2 D^{b}\left(p_{a b}\right)+\pi \partial_{a} \phi\right).
	\end{aligned}
\end{equation}
The consistency condition of the two primary constraints gives, respectively, 
\begin{equation}
\begin{aligned}
\label{Hamiltonianconstraint}
\dot{M}&=-\frac{\delta H_{t}}{\delta N}=:H\\&=\frac{2}{\sqrt{h}}\left(\frac{p_{a b} p^{a b}-\frac{1}{2} p^{2}}{\phi}+\frac{(p-\phi \pi)^{2}}{2 \phi(3+2 \omega)}\right) \\
&\!+\! \frac{1}{2} \sqrt{h}\left(-\phi R+\frac{\omega(\phi)}{\phi}\left(D_{a} \phi\right) D^{a} \phi+2 D_{a} D^{a} \phi+2 \xi(\phi)\right)\\
&=0
\end{aligned}
\end{equation}
and
\begin{eqnarray}
\label{vectorconstraint}
\dot{M_{a}}\!=\!-\frac{\delta H_{t}}{\delta N^{a}}=:V_{a}\!=\!-2 D^{b}\left(p_{a b}\right)+\pi \partial_{a} \phi=0.
\end{eqnarray}
It turns out that, by considering the compact $\Sigma$ without a boundary, the total Hamiltonian consists of two secondary constraints as
\begin{equation}
\label{totalhamiltonian}
H_{t}=\int d^{3} x\left(N^{a} V_{a}+N H\right).
\end{equation}
The phase space is equipped with the symplectic structure defined by
\begin{eqnarray}
\begin{aligned}
\label{symplectic structure}
\left\{h_{a b}(x), p^{c d}(y)\right\} &=\delta_{a}^{(c} \delta_{b}^{d)} \delta^{3}(x, y) , \\ \{\phi(x), \pi(y)\} &=\delta^{3}(x, y) .
\end{aligned}
\end{eqnarray}
Straightforward calculations show that the constraints comprise a first-class system\cite{zhang2011nonperturbative}.

In the sector of $\omega(\phi) =-\frac{3}{2}$, following the same as above, one obtains the total Hamiltonian for the compact $\Sigma$ without a boundary as\cite{olmo2011hamiltonian}
\begin{equation}
\label{totalhamiltonian2}
H^{\prime}_{t}=\int_{\Sigma} d^{3} x\left(N^{a} V_{a}+N H^{\prime}+\lambda S\right),
\end{equation}
where $V_{a}$ is given by Eq.~(\ref{vectorconstraint}), the smeared Hamiltonian constraint reads
\begin{multline}
	\label{Hamiltonian constraint2}
	H^{\prime}(N) \!=\! \int d^{3} x\, N H^{\prime} 
	= \int d^{3} x\, N\left[\frac{2}{\sqrt{h}}\left(\frac{p_{a b} p^{a b}-\frac{1}{2} p^{2}}{\phi}\right)\right. \\
	\left. + \frac{1}{2} \sqrt{h}\left(-\phi R-\frac{3}{2 \phi}\left(D_{a} \phi\right) D^{a} \phi+2 D_{a} D^{a} \phi+2 \xi(\phi)\right)\right],
\end{multline}
and the smeared conformal constraint reads 
\begin{equation}
\label{conformal constraint}
S(\lambda)=\int d^{3} x \lambda S=\int d^{3} x \lambda(p-\pi \phi).
\end{equation}
The gauge transformations generated by $S(\lambda)$ can be calculated as
\begin{equation}
\begin{aligned} 
\label{conformaltransformation}
\left\{h_{a b}, S(\lambda)\right\}&=\lambda h_{a b}, \ \left\{p^{a b}, S(\lambda)\right\}=-\lambda p^{a b}, \\
\{\phi, S(\lambda)\}&=-\lambda \phi, \ \{\pi, S(\lambda)\}=\lambda \pi.
\end{aligned}
\end{equation}
They correspond to the following transformations:
\begin{equation}
\begin{aligned} 
\label{conformaltransformation2}
g_{\mu \nu} \rightarrow e^\lambda g_{\mu \nu}, \quad \phi \rightarrow e^{-\lambda} \phi.
\end{aligned}
\end{equation}
Hence, $S(\lambda)$ is called as a conformal constraint. The consistency condition of the conformal constraint leads to the condition\cite{zhang2011nonperturbative}
\begin{equation}
\label{secondary constraint}
2 \xi(\phi)-\phi \xi^{\prime}(\phi)=0.
\end{equation}
For the vacuum case that we considered, the solutions to Eq.~(\ref{secondary constraint}) are either $\xi(\phi)=0$ or $\xi(\phi)=C \phi^{2}$, where $C$ is a certain dimensional constant. Thus, the consistency condition strongly restricted the feasible scalar-tensor theory in the sector $\omega(\phi) =-\frac{3}{2}$ to only two possible theories. In either of the theories, the three kinds of constraints comprise a first-class constraint system.

\subsection{The connection-dynamical formalism}
To obtain the connection-dynamical formalism of the scalar-tensor theory, one first introduces the transformation
\begin{equation}
\label{tildeK}
\tilde{K}^{a b}=\phi K^{a b}+\frac{h^{a b}}{2 N}\left(\dot{\phi}-N^{c} \partial_{c}. \phi\right)
\end{equation}
Then, one defines a $\mathrm{su(2)}$-valued one-form field as $\tilde{K}_a^i=\tilde{K}_{a b} e^{b i}$ and a densitized triad field as $E_{i}^{a}=\sqrt{h} e_{i}^{a}$, where $e_{i}^{a}$ is the triad field that satisfies $h_{a b}=\delta_{i j} e_a^i e_b^j$, with $i,j=1,2,3$ being the indices of internal $\mathrm{su(2)}$ Lie algebra, raised and lowered by $\delta_{i j}$. The fields $\tilde{K}_{a}^{i}$ and $E_{j}^{b}$ satisfy a new constraint, known as the Gaussian constraint\cite{thiemann2008modern}:
\begin{equation}
\label{Gaussian constraint1}
G_i=\epsilon_{i j}{ }^k \tilde{K}_a^j E_k^a=0.
\end{equation}                                     
The covariant derivative can be extended to the internal space as
\begin{equation}
\label{expand covariant derivative}
D_a v^i=\partial_a v^i+\omega_a{ }^i{ }_j v^j,
\end{equation}
where the spin connection $\omega_a{ }^i{ }_j$ is determined by compatibility condition $D_{a}e_{b}^{i}=0$. The dual version of the spin connection reads $\Gamma_a^i=-\frac{1}{2} \epsilon^{i j k} \omega_{a j k}$, which can be expressed as 
\begin{equation}
\begin{aligned}
\label{spin connection}
 \Gamma_{a}^{i} &=\frac{1}{2} \epsilon^{i j k} e_{k}^{b}\left(\partial_{b} e_{a}^{j}-\partial_{a} e_{b}^{j}+e_{j}^{c} e_{a}^{l} \partial_{b} e_{c}^{l}\right) \\ &\!=\! \frac{1}{2} \epsilon^{i j k} E_{k}^{b}\left(\partial_{b} E_{a}^{j}-\partial_{a} E_{b}^{j}+E_{j}^{c} E_{a}^{l} \partial_{b} E_{c}^{l}+E_{a}^{j} \frac{\partial_{b} \operatorname{det} E}{\operatorname{det} E}\right) ,
\end{aligned}
\end{equation}
where $\operatorname{det} E$ is the determinant of $E_{i}^{a}$ satisfying $|\operatorname{det} E|=h$. Now the new canonical geometric variables can be defined as
\begin{equation}
\label{new variables}
A_{a}^{i}=\Gamma_{a}^{i}+\gamma \tilde{K}_{a}^{i}, \quad E_{i}^{a}=\sqrt{h} e_{i}^{a},
\end{equation}
where $\gamma$ is the Barbero-Immirzi parameter, which is real valued. The Gaussian constraint, given in Eq.~(\ref{Gaussian constraint1}), can be reformulated in terms of the new canonical geometric variables as
\begin{equation}
\label{Gaussian constraint2}
\mathcal{G}_{i}=\mathscr{D}_{a} E_{i}^{a} \equiv \partial_{a} E_{i}^{a}+\epsilon_{i j k} A_{a}^{j} E^{a}_{k}.
\end{equation}  
Now the nonzero Poisson bracket between the new canonical geometric variables reads
\begin{equation}
\begin{aligned}
\label{Poisson brackets}
\left\{A_{a}^{j}(x), E_{k}^{b}(y)\right\} &=\gamma \delta_{a}^{b} \delta_{k}^{j} \delta(x, y). 
\end{aligned}
\end{equation}
It can be proved that Eq.~(\ref{Poisson brackets}) is equivalent to the Poisson brackets in Eq.~(\ref{symplectic structure}) when the Gaussian constraint is imposed\cite{zhang2011nonperturbative}.

In the sector of $\omega(\phi) \neq-\frac{3}{2}$,
%By substituting the new canonical variables, the total Hamiltonian consists of three constraints
%\begin{equation}
%\label{total Hamiltonian1}
%H_{t}=\int d^{3}x \Lambda^{i} \mathcal{G}_{i}+N^{a} V_{a}+N H
%\end{equation}
the Gaussian constraint has been expressed as Eq.~(\ref{Gaussian constraint2}). The diffeomorphism and Hamiltonian constraints can, respectively, be expressed in terms of the new variables, up to the Gaussian constraint, as 
\begin{equation}
\label{vectorconstraint2}
V_{a}=\frac{1}{\gamma} F_{a b}^{i} E_{i}^{b}+\pi \partial_{a} \phi
\end{equation}
and
\begin{equation}
\begin{aligned}
\label{Hamiltonianconstraint2} 
H &=\frac{\phi}{2}\left[F_{a b}^{j}-\left(\gamma^{2}+\frac{1}{\phi^{2}}\right) \varepsilon_{j m n} \tilde{K}_{a}^{m} \tilde{K}_{b}^{n}\right] \frac{\varepsilon_{j k l} E_{k}^{a} E_{l}^{b}}{\sqrt{h}} \\ &+\frac{1}{3+2 \omega(\phi)}\left(\frac{\left(\tilde{K}_{a}^{i} E_{i}^{a}\right)^{2}}{\phi \sqrt{h}}+2 \frac{\left(\tilde{K}_{a}^{i} E_{i}^{a}\right) \pi}{\sqrt{h}}+\frac{\pi^{2} \phi}{\sqrt{h}}\right) \\ &+\frac{\omega(\phi)}{2 \phi} \sqrt{h}\left(D_{a} \phi\right) D^{a} \phi+\sqrt{h} D_{a} D^{a} \phi+\sqrt{h} \xi(\phi),
\end{aligned}
\end{equation}
where $F_{a b}^i \equiv 2 \partial_{[a} A_{b]}^i+\epsilon_{k l}^i A_a^k A_b^l$ is the curvature of $A_{a}^{i}$. Straightforward calculations show that the constraints still comprise a first-class system\cite{zhang2011nonperturbative}. The Hamiltonian is still the linear combination of the three constraints.

In the sector of $\omega(\phi) =-\frac{3}{2}$, the total Hamiltonian is the linear combination of four constraints:
\begin{equation}
\label{total Hamiltonian2}
H^{\prime}_{t}=\int d^{3}x (\Lambda^{i} \mathcal{G}_{i}+N^{a} V_{a}+N H^{\prime}+\lambda S),
\end{equation}
where the Gaussian constraint and diffeomorphism constraint are the same as in the case of $\omega(\phi) \neq-\frac{3}{2}$ and the Hamiltonian and conformal constraints can, respectively, be expressed as 
\begin{equation}
\begin{aligned} 
\label{Hamiltonianconstraint3} 
H^{\prime} &=\frac{\phi}{2}\left[F_{a b}^{j}-\left(\gamma^{2}+\frac{1}{\phi^{2}}\right) \varepsilon_{j m n} \tilde{K}_{a}^{m} \tilde{K}_{b}^{n}\right] \frac{\varepsilon_{j k l} E_{k}^{a} E_{l}^{b}}{\sqrt{h}} \\ &-\frac{3}{4 \phi} \sqrt{h}\left(D_{a} \phi\right) D^{a} \phi+\sqrt{h} D_{a} D^{a} \phi+\sqrt{h} \xi(\phi) \\ 
\end{aligned}
\end{equation}
and
\begin{equation}
\begin{aligned} 
\label{conformal constraint2}
S &=-\tilde{K}_{a}^{i} E_{i}^{a}-\pi \phi .
\end{aligned}
\end{equation}
Straightforward calculations show that the constraints still comprise a first-class system\cite{zhang2011nonperturbative}.

\section{Deparametrization and Loop Quantization}
\label{dlq}

As shown in the Hamiltonian analysis, the scalar-tensor theory is a first-class constraint system, similar to GR. In such a system, the total Hamiltonian generates gauge transformations, such that the Dirac observables satisfy $\dot{O}=\{O,H_{t}\}=0$. So, they appear not to evolve with time\cite{thiemann2008modern}. To address this issue, the concept of relative evolution was introduced\cite{rovelli1991observable,rovelli1991quantum,rovelli1990quantum,thiemann2006solving}. Deparametrization is a method for implementing relative evolution by designating one degree of freedom in a theory as time and studying the dynamical evolution of the other degrees of freedom relative to it. In principle, deparametrization can also be applied to GR to analyze the evolution of physical quantities relative to intrinsic distances, angles, etc. However, the calculations tend to be overly complex and the results are unclear\cite{bodendorfer2015general,duch2014observables,duch2015addendum}. In the scalar-tensor theory, there exists a scalar field nonminimally coupled with the geometric part. Thus, it is interesting to study whether the scalar field can be taken as time to understand the evolution of the geometric part relative to it.

In the classical theory of a completely constrained system with finite degrees of freedom, if the Hamiltonian constraint took the form $C=\pi - h$, where $\pi$ is the momentum conjugate to the scalar $\phi$, the Poisson commutativity of a Dirac observable $O$ with $C$ indicates $\frac{\partial}{\partial \phi} O=\{O,\ h\}$. This means that the equations of motion for Dirac variables describe their evolution relative to $\phi$, with $h$ serving as a physical Hamiltonian. In the quantum theory, a Hamiltonian constraint of this type yields a quantum evolution equation similar to the Schr\"{o}dinger equation, given by $i \frac{d}{d \phi} |\Psi \rangle =\hat{h} |\Psi \rangle$. Thus, to deparametrize the scalar-tensor theory by treating the scalar field $\phi$ as time, we need to solve for its conjugate momentum from the Hamiltonian constraint. The constraint surface generated by the new constraint of the form $C=\pi-h$ should be identical to the original one. Moreover, the gauge transformations generated by the new constraint differs from the original ones only in the gauge parameter\cite{thiemann2008modern}. Consequently, the dynamical evolution relative to $\phi$ is governed by the physical Hamiltonian $h$.

\subsection{Deparametrization of the scalar-tensor theory}    
In the sector of $\omega(\phi) \neq-\frac{3}{2}$, by Eq.~(\ref{Hamiltonianconstraint2}), the Hamiltonian constraint $H=0$ can be solved as
\begin{align}
\label{solve pi} 
\pi=-\frac{\tilde{K}_{a}^{i} E_{i}^{a}}{\phi} \pm \sqrt{t(\phi,A,E)},
\end{align} 
where the sign $\pm$ defines the following different regions in the phase space: The $+$ sign corresponds to the region $\Gamma_{+}$ satisfying $\pi+\frac{\tilde{K}_{a}^{i} E_{i}^{a}}{\phi} \geq 0$, and the $-$ sign corresponds to the region $\Gamma_{-}$ satisfying $\pi+\frac{\tilde{K}_{a}^{i} E_{i}^{a}}{\phi} \leq 0$. Here, the function $t(\phi,A,E)$ is given by
\begin{align}
\label{t} 
&t(\phi,A,E) \nonumber\\
=&-\frac{3+2\omega(\phi)}{2}\left[F^{j}_{a b}-(\gamma^{2}+\frac{1}{\phi^{2}}) \epsilon_{j m n} \tilde{K^{m}_{a}} \tilde{K^{n}_{b}} \right] \epsilon_{j k l}E^{a}_{k}E^{b}_{l} \nonumber\\&-\frac{(3+2\omega(\phi))\omega(\phi)h}{2\phi^{2}}(D_{a}\phi)D^{a}\phi \nonumber \\&-\frac{(3+2\omega(\phi))h}{\phi}(D^{a}D_{a}\phi+\xi(\phi)).
\end{align}
By substituting the function $t(\phi, A, E)$ into Eq.~(II.29) and imposing the Hamiltonian constraint $H=0$, one finds that the function $t(\phi, A, E)$ satisfies
\begin{equation}
	t(\phi,A,E)
	=
	\left(\pi+\frac{\tilde{K}_{a}^{i} E_{i}^{a}}{\phi}\right)^2 ,
\end{equation}
which implies $t(\phi, A, E) \geq 0$ on the constraint surface. Hence, the Hamiltonian constraint can be reformulated as
\begin{align}
\label{new constraint1} 
C(x)=\pi(x)-\left(-\frac{\tilde{K}_{a}^{i} E_{i}^{a}}{\phi} \pm \sqrt{t(\phi,A,E)}\right)
\end{align}    
The deparametrized system is composed of the Gaussian constraint (\ref{Gaussian constraint2}), the diffeomorphism constraint (\ref{vectorconstraint2}), and the new Hamiltonian constraint (\ref{new constraint1}), which is equivalent to the original constraint system. It remains a first-class constraint system\cite{thiemann2008modern}.

In the sector of $\omega(\phi) =-\frac{3}{2}$, the Hamiltonian constraint (\ref{Hamiltonianconstraint3}) does not contain $\pi$. We first solve the conformal constraint $S=0$ by Eq.~(\ref{conformal constraint}) to obtain $p=\pi\phi$. Then, we insert this result into the Hamiltonian constraint (\ref{Hamiltonian constraint2}) and solve $H'=0$ for $\pi$. The result reads
\begin{align}
\label{solve pi1} 
\pi \!=\! \pm \! \sqrt{\frac{2p_{a b}p^{a b}}{\phi^{2}}\!-\! \frac{hR}{2}\!-\! \frac{3h}{4\phi^{2}}\!(D_{a}\phi)\! D^{a}\phi \!+\! \frac{h}{\phi}D_{a}D^{a}\phi \!+\! \frac{h \xi(\phi)}{\phi}},
\end{align}
which can be expressed in terms of the connection variables as
\begin{align}
\label{solve pi2}
\pi = \pm \sqrt{t^{\prime}(\phi,A,E)},
\end{align} 
where the $+$ sign corresponds to the region $\Gamma^{\prime}_{+}$ satisfying $\pi \geq 0$  and the $-$ sign corresponds to the region $\Gamma^{\prime}_{-}$ satisfying $\pi \leq 0$. Here, the function $t^{\prime}(\phi,A,E)$ is given by   
\begin{align}
\label{tprime}  
& t^{\prime}(\phi,A,E) \nonumber \\
= &\left[\epsilon_{j k l}F^{j}_{a b}+(\gamma^{2}+\frac{1}{\phi^{2}})\tilde{K}^{l}_{a}\tilde{K}^{k}_{b}+(\frac{1}{\phi^{2}}-\gamma^{2})\tilde{K}^{k}_{a}\tilde{K}^{l}_{b}\right]E^{a}_{k}E^{b}_{l} \nonumber\\
&-\frac{3h}{4\phi^{2}}(D_{a}\phi)D^{a}\phi+\frac{h}{\phi}D_{a}D^{a}\phi +\frac{h \xi(\phi)}{\phi},
\end{align}       
which satisfies $t^{\prime}(\phi,A,E) \geq 0$ on the constraint surface. Hence, the Hamiltonian constraint can be reformulated as
\begin{align}
\label{new constraint2} 
C^{\prime}(x)=\pi(x)-\left(\pm \sqrt{t^{\prime}(\phi,A,E)}\right).
\end{align}    
The deparametrized system is composed of the Gaussian constraint (\ref{Gaussian constraint2}), the diffeomorphism constraint (\ref{vectorconstraint2}), the conformal constraint (\ref{conformal constraint2}), and the new Hamiltonian constraint E(\ref{new constraint2}), which is equivalent to the original constraint system. 

\subsection{Loop quantization of the scalar-tensor theory} 

Since the phase space of the scalar-tensor theory consists of the geometric part($A_{a}^{i},E_{i}^{a}$) and the scalar field part($\phi,\pi$), it is natural to construct the total kinematical Hilbert space of the quantum theory as the tensor product of the Hilbert spaces of the geometric part and the scalar field part. We define the Hilbert space of the geometric part to be the same as in the usual LQG. The smooth connection configuration space is extended to the quantum connection configuration space, denoted by $\bar{\mathcal{A}}$. Given an edge $e \subset \Sigma$, its associated quantum configuration space $\overline{\mathcal{A}}_{e}$ is isomorphism to the group $\mathrm{SU(2)}$ via the map 
\begin{align}
\label{holonomy} 
h_e(A)=\mathcal{P} \exp \{\int_e A_{a}^{i} \tau_{i}\},
\end{align}  
where $\mathcal{P}$ is the path-ordering operator along the edge $e$ and $\tau_{i}=-i \frac{\sigma_{i}}{2}$, with $\sigma_{i}$ being the Pauli matrices, are the generators of the Lie algebra $\mathrm{su(2)}$. Through the projective technique, $\bar{\mathcal{A}}$ is naturally equipped with the Ashtekar-Lewandowski measure, denoted by $\mu_{AL}$\cite{ashtekar2004background,han2007fundamental}. Thus, the kinematical Hilbert space of geometric part is obtained straightforwardly as $\mathcal{H}_{\mathrm{k i n}}^{\mathrm{g e o}}=L^{2}(\bar{\mathcal{A}},d\mu_{AL})$, which can be viewed as the completion of the cylindrical functions with respect to the inner product, i.e., $\mathcal{H}_{\mathrm{k i n}}^{\mathrm{g e o}}= \overline{Cyl^{\mathrm{g e o}}}$. Note that the space of cylindrical functions can be written as the following finite linear combination:
\begin{align}
\label{cylindrical functions1} 
Cyl^{\mathrm{g e o}} :=Span \{ |\beta, \vec{j}, \vec{\iota}\rangle \},
\end{align} 
where  
\begin{align}
\label{basis1} 
|\beta, \vec{j}, \vec{\iota}\rangle[A]=&\left(\prod_{v \in V(\beta)}\left(\iota_v\right)_{m_1^v \cdots m_{N_v}^v}^{n_1^v \cdots n^v_{N^{\prime}_v}}\right) \nonumber\\
&\times \left(\prod_{e \in E(\beta)} \sqrt{2 j_e +1} D^{\left(j_e\right) m_e}{ }_{n_e}\left(h_e\right)\right)
\end{align} 
is an orthonormal basis in $\mathcal{H}_{\mathrm{k i n}}^{\mathrm{g e o}}$. Here, $\beta$ is a graph with edges $E(\beta)=\{e_{1}, \cdots, e_{n}\}$ and vertices $V(\beta)=\{v_{1}, \cdots, v_{k}\}$, $\iota_v$ is an arbitrary tensor, $j_{e}$ is a nontrivial irreducible representation of $\mathrm{SU(2)}$, and $D^{\left(j_e\right) m_e}{ }_{n_e}\left(h_e\right)$ is the Wigner-D matrix\cite{makinen2019dynamics}. For two cylindrical functions $\psi$ and $\psi^{\prime}$, defined over graphs $\beta$ and $\beta^{\prime}$, respectively, one finds a larger graph $\beta^{\prime \prime} =\left\{e_1^{\prime \prime}, \ldots, e_{n^{\prime \prime}}^{\prime \prime}\right\}$ such that both $\psi$ and $\psi^{\prime}$ can be regarded as cylindrical functions on it. Then, their inner product is given by
\begin{align}
\label{inner product1} 
\left\langle\psi \mid \psi^{\prime}\right\rangle=\int d g_1 \ldots d g_{n^{\prime \prime}} \bar{\psi}\left(g_1, \ldots, g_{n^{\prime \prime}}\right) \psi^{\prime}\left(g_1, \ldots, g_{n^{\prime \prime}}\right).
\end{align} 
The fluxes of a densitized triad on 2-surfaces $\mathrm{S}$ are defined by
\begin{align}
\label{flux} 
E(\mathrm{S}, \mathrm{g}):=\int_\mathrm{S} \epsilon_{a b c} E_i^a \mathrm{g}^i,
\end{align}  
where $\epsilon_{a b c}$ is the Levi-Civita tensor density of weight $-1$ on $\Sigma$ and $\mathrm{g}^{i}$ is a su(2)-valued function on $\mathrm{S}$. The fundamental operators of geometry part act on any cylindrical function $\psi(A)$ as
\begin{equation}
\begin{aligned}
\label{operator1} 
\hat{h}_e(A) \cdot \psi(A)&=h_e(A) \psi(A), \\  \hat{E}(\mathrm{S}, \mathrm{g}) \cdot \psi(A)&=i\{E(\mathrm{S}, \mathrm{g}), \psi(A)\}.
\end{aligned}
\end{equation} 
The geometric operators constructed in LQG, such as the area operator\cite{rovelli1995discreteness}, the volume operator\cite{ashtekar1997quantum}, and the length operator\cite{thiemann1998length,ma2010new}, are still valid in $\mathcal{H}_{\mathrm{k i n}}^{\mathrm{g e o}}$.

We now introduce the polymerlike  representation for the scalar field part\cite{lewandowski2016loop}. A quantum state $|\phi \rangle$ is labeled by a function
\begin{align}
\label{basis2} 
\phi: \Sigma \rightarrow \mathbb{R},
\end{align} 
where the differentiability class $C^{k}$ is to be fixed depending on applications. The space of cylindrical functions of scalar field part, denoted by $Cyl^{\mathrm{sc}}$, is defined by the following finite linear combination: 
\begin{align}
\label{cylindrical functions2} 
Cyl^{\mathrm{sc}} :=Span \{ |\phi \rangle \}.
\end{align} 
The inner product in this space is defined as
\begin{align}
\label{inner product2} 
\langle\phi \mid \phi\rangle=1, \quad \mathrm{and} \quad\left\langle\phi \mid \phi^{\prime}\right\rangle=0  \quad \mathrm{whenever} \quad \phi \neq \phi^{\prime}.
\end{align} 
The kinematic Hilbert space $\mathcal{H}_{\mathrm{k i n}}^{\mathrm{sc}}$ is the completion of the cylindrical functions with respect to the above inner product, i.e., $\mathcal{H}_{\mathrm{k i n}}^{\mathrm{sc}}= \overline{Cyl^{\mathrm{sc}}}$. In the momentum representation, the kinematic Hilbert space can be viewed as being spanned by the functions
\begin{align}
\label{basis21} 
|\phi\rangle[\pi]=\exp \left(-i \int \mathrm{d}^3 x \phi(x) \pi(x)\right).
\end{align} 
The scalar field operator $\hat{\phi}(x)$ acts on the basis state as
\begin{align}
\label{operator2} 
\hat{\phi}(x) \cdot|\phi\rangle=\phi(x)|\phi\rangle,
\end{align} 
while the momentum operator acts it as
\begin{align}
\label{operator3}
\exp \left(-i \int \mathrm{d}^3 x \mathrm{f}(x) \hat{\pi}(x)\right)\cdot|\phi\rangle=|\phi+\mathrm{f}\rangle,
\end{align} 
where $\mathrm{f}$ is a function on $\Sigma$. In this representation, the derivatives of scalar field can be directly defined as operators by
\begin{align}
\label{operator4}
\widehat{\partial_a \ldots \partial_b \phi}(x) \cdot |\phi\rangle=\partial_a \ldots \partial_b \phi(x)|\phi\rangle.
\end{align} 
The total kinematic Hilbert space of the scalar-tensor theory is constructed as the tensor product of the geometric part and the scalar field part as
\begin{align}
\label{the total kinematic Hilbert space}
\mathcal{H}_{\mathrm{kin} }^{t}= \mathcal{H}_{\mathrm{kin} }^{\mathrm{g e o} } \otimes \mathcal{H}_{\mathrm{kin} }^{\mathrm{s c}}. 
\end{align} 

The Gaussian constraint can be quantized directly as a well-defined operator in $\mathcal{H}_{\mathrm{k i n}}^{t}$, analogous to its treatment in the usual LQG. Its kernel forms the $\mathrm{SU(2)}$ gauge invariant Hilbert space $\mathcal{H}_{G}^{t}$. The basis of $\mathcal{H}_{G}^{t}$ consists of states $|\beta, \vec{j}, \vec{\iota}\rangle \otimes |\phi \rangle$, where the tensors $\vec{\iota}$ assigned to the vertices of graph $\beta$ are the intertwiners that are invariant under the internal gauge transformations.

Since the spatial diffeomorphism constraint generates the spatial diffeomorphism transformations, one can use the group-averaging method to solve this constraint\cite{ashtekar1995quantization}. Each spatial diffeomorphism $f \in Diff$ defines an unitary operator as
\begin{align}
\label{diffeomorphism operator}
\hat{U}_{f} \cdot|\beta, \vec{j}, \vec{\iota}\rangle \otimes |\phi \rangle=|f(\beta), \vec{j} \circ f^{-1}, \vec{\iota} \circ f^{-1} \rangle \otimes |\phi \circ f^{-1} \rangle.
\end{align} 
We first define a projection map acting on cylindrical functions as
\begin{align}
\label{projection map}
\hat{P}_{Diff_{\beta}} \cdot|\Psi_\beta \rangle:=\frac{1}{n_\beta} \sum_{g \in G S_\beta} \hat{U}_{g} \cdot|\Psi_\beta \rangle,
\end{align} 
where $|\Psi_\beta \rangle \in Cyl_{G}^{\mathrm{g e o}} \otimes Cyl^{\mathrm{sc}}=:Cyl_{G}$ is a state in the space of gauge invariant cylindrical functions, $G S_\beta=Diff_\beta / T Diff_\beta$ represents the group of graph symmetries, with $Diff_{\beta}$ being the group of all diffeomorphisms preserving the colored graph $\beta$ and $T Diff_\beta$ being the
group of diffeomorphisms which trivially act on $\beta$, and $n_{\beta}$ denotes the number of the elements in $G S_\beta$. Then, we define the rigging map $\eta: Cyl_{G} \rightarrow Cyl^{\star}$(the algebraic dual space of $Cyl_{G}$) by the action 
\begin{align}
\label{rigging map}
\eta\left(\Psi_\beta \right)\left[\Phi_{\beta^{\prime}}\right]:=\sum_{f \in Diff / D i f f_\beta}\left\langle\hat{U}_{f} \hat{P}_{D i f f_\beta} \Psi_\beta \mid \Phi_{\beta^{\prime}}\right\rangle_{kin}.
\end{align}
Furthermore, we can define the inner product on $\eta[Cyl_{G}]$ as 
\begin{align}
\label{Dinner product}
\left\langle\eta\left(\Psi_\beta\right) \mid \eta\left(\Phi_{\beta^{\prime}}\right)\right\rangle_{Diff}:=\eta\left(\Psi_\beta\right)\left[\Phi_{\beta^{\prime}}\right].
\end{align}
The diffeomorphism invariant Hilbert space $\mathcal{H}_{Diff}$ is defined by the completion of $\eta[Cyl_{G}]$ with  respect to the inner product given by Eq.~(\ref{Dinner product}).

\subsection{Quantum dynamics of the scalar-tensor theory}
In the sector of $\omega(\phi) \neq -\frac{3}{2}$, to quantize the deparametrized Hamiltonian constraint (\ref{new constraint1}), we divide its smearing version into the following three terms:
\begin{equation}
\begin{aligned}
\label{three terms}
&C_1(N)=\int d^3 x N(x) \pi(x), \\ & C_2(N)=\int d^3 x N(x) \frac{\tilde{K}_a^i E_i^a}{\phi}(x), \\  &C_3(N)=\mp \int d^3 x N(x) \sqrt{t(\phi,A,E)}
\end{aligned}
\end{equation}
and then quantize them respectively. 

For the first term $C_{1}(N)$, by introducing a small dimensionless constant $\lambda_0$, its corresponding operator can be defined as
\begin{align}
	\begin{split}
		\label{term1}
		\widehat{C}_1(N) =& \frac{1}{2 i \lambda_0} \biggl(  \exp \left[i \lambda_0 \int d^3 x N(x) \hat{\pi}(x)\right] \\
		& - \exp \left[-i \lambda_0 \int d^3 x N(x) \hat{\pi}(x)\right] \biggr).
	\end{split}
\end{align}

For the second term $C_2(N)$, one needs to define the inverse operator $\hat{\phi}^{-1}(x)$, because 0 is in the point spectrum of operator $\hat{\phi}(x)$. By using the classical identity
\begin{align}
\label{phi-1/2}
\frac{1}{\sqrt{|\phi(x)|}}=&\frac{2 \mathrm{sgn}(\phi(x))}{i \lambda_0}\left\{\sqrt{|\phi(x)|}, \exp \left(i \lambda_0 \int d^3 y \pi(y)\right)\right\} \nonumber \\ 
\times &\exp \left(-i \lambda_0 \int d^3 y \pi(y)\right),
\end{align}
we can define an operator acting on the basis as
\begin{align}
\label{phi-1/2operator}
&\widehat{ \frac{1}{\sqrt{|\phi(x)|}} }\cdot|\phi\rangle \nonumber=\\&\frac{-2 \mathrm{sgn}(\phi(x))}{\lambda_0} \left[\sqrt{|\hat{\phi}(x)|},\exp \left(i \lambda_0 \int d^3 y \hat{\pi}(y)\right)\right]\nonumber\\  &\cdot \exp \left(-i \lambda_0 \int d^3 y \hat{\pi}(y)\right)\cdot |\phi\rangle \nonumber\\
&=\frac{-2 \mathrm{sgn}(\phi(x))}{\lambda_0}\left(\sqrt{|\phi(x)|}-\sqrt{|\phi(x)+\lambda_0|}\right)|\phi\rangle,
\end{align}
where $\mathrm{sgn}(\phi(x))$, taking values $+1$ or $-1$, is the symbolic function. Then, the operator $\hat{\phi}^{-1}(x)$ can be defined by acting twice as
\begin{align}
\label{phi-1operator}
\hat{\phi}^{-1}(x) \cdot |\phi\rangle&\!=\! \frac{4 \mathrm{sgn}(\phi(x))}{\lambda_0{ }^2}\! \left(\sqrt{|\phi(x)|}-\sqrt{|\phi(x)+\lambda_0|} \! \right)^2|\phi\rangle \nonumber \\ &\equiv D(x)|\phi\rangle.
\end{align}
Moreover, we introduce the following classical identities to regularize $C_2(N)$:
\begin{align}
	\label{E}
	E_i^a&=\frac{1}{2} \varepsilon^{a b c} \varepsilon_{i j k} e_b^j e_c^k \operatorname{sgn}\left(\operatorname{det}\left(e_a^i\right)\right), \\
	\label{sgne}
	\operatorname{sgn}\left(\operatorname{det}\left(e_a^i\right)\right)&=\frac{1}{6} \frac{1}{\sqrt{h}} \varepsilon_{i j k} \varepsilon^{a b c} e_a^i e_b^j e_c^k, \\
	\label{e}
	e_a^i&=\frac{2}{\gamma}\left\{A_a^i, V\right\}, \\
	\label{K}
	\tilde{K}_a^i&=\gamma^{-\frac{5}{2}}\left\{A_a^i,\left\{H^E(1), V\right\}\right\},
\end{align}
where 
\begin{align}
\label{HE1}
H^E(1)=\frac{1}{2} \int d^3 x F_{a b}^j \frac{\varepsilon_{j k l} E_k^a E_l^b}{\sqrt{h}}.
\end{align}
Using above identities and the point-splitting scheme, the second term can be expressed as 
\begin{align} 
\label{C2}
& C_2(N)=\int d^3 x N(x) \frac{\tilde{K}_a^i E_i^a}{\phi}(x) \nonumber\\ & =18 \gamma^{-\frac{15}{2}} \lim _{\epsilon \rightarrow 0} \! \int \! d^3 x N(x) \phi^{-1}(x) \! \left\{ \! A_a^i(x),\! \left\{ \! H^E(1), V_{(x,\epsilon)} \! \right\} \! \right\} \nonumber\\ &\times \varepsilon^{a b c} \varepsilon_{i j k}\left\{A_b^j(x), V_{(x,\epsilon)}\right\}\left\{A_c^k(x), V_{(x,\epsilon)}\right\} \nonumber\\ & \! \times \! \int \! d^3 y \chi_{\epsilon}(x-y) \varepsilon_{l m n} \varepsilon^{d e f} \! \left\{A_l^d(y), V_{(y,\epsilon)}^{\frac{2}{3}} \! \right\} \! \left\{A_e^m(y), V_{(y,\epsilon)}^{\frac{2}{3}} \! \right\} \nonumber\\ &\times \left\{A_f^n(y), V_{(y,\epsilon)}^{\frac{2}{3}}\right\},
\end{align}
where $\chi_{\epsilon}(x-y)$ is the characteristic function such that $\lim _{\epsilon \rightarrow 0} \frac{\chi_\epsilon(x-y)}{\epsilon}=\delta(x-y)$ and $V_{(y, \epsilon)}=\int d^3 z \chi_\epsilon(y-z) \sqrt{h}(z)$ represents the volume of the 
neighborhood $(y, \epsilon)$ around $y$. To express $C_{2}(N)$ in terms of holonomies and other variables with clear quantum analogs, we choose the triangulation that depends on the graphs\cite{han2007fundamental,thiemann2008modern,zhang2011extension}. Given a graph $\beta$, by selecting three edges for each vertex $v \in \beta(v)$, one can obtain a tetrahedron $\Delta_{\beta, v, e_l, e_j, e_K}^\delta$ spanned by three segments $\left(s_I, s_J, s_K\right)$, where each segment is part of edge with coordinate length $\delta$. For each $\Delta_{\beta, v, e_l, e_j, e_K}^\delta$, one can construct seven additional tetrahedrons by backward analytic extension of the segments. The regions of $\Sigma$ without a vertex of $\beta$ can be triangulated arbitrarily. Then, we have the approximations
\begin{align}
	\label{h}
	h_{s_I} &\approx 1-\delta \dot{s}_I^a A_a^i \tau_i, \\
	\label{epsilon}
	\int d^3 x \varepsilon^{a b c} &\approx \sum_{\Delta} \frac{1}{6} \delta^3 \epsilon\left(s_I s_J s_K\right) \varepsilon^{I J K} \dot{s}_I^a \dot{s}_J^b \dot{s}_K^c,
\end{align}
where $h_{s_I}$ is the holonomy along $s_I$, $\dot{s}_I^a$ is the tangent vector of $s_I$, and $\epsilon\left(s_I s_J s_K\right)=\operatorname{sgn}\left(\operatorname{det}\left(\dot{s}_I \dot{s}_J \dot{s}_K\right)\right)$ takes the values
$+1,-1,0$ if the tangents of the three segments $\left( s_I, s_J, s_K \right)$ at $v$ (in that sequence) form a matrix of
positive, negative, or vanishing determinant. Using these approximations, we can regularize $C_2(N)$ as
\begin{align} 
	\label{C2delta}
	& C_2^{\delta}\! ( \! N \!)\!=\!2^9 \! \gamma^{-\frac{15}{2}} \! \lim _{\epsilon \rightarrow 0} \! \sum_{v \in \beta(v)} \! \frac{N \!(\! v \!) \phi^{-1} \!(\! v \!)}{E \! (\! v \!)\!} \! \sum_{v(\Delta)=v} \! \epsilon \!\left(\! s_I s_J s_K \! \right) \! \varepsilon^{L J K} \nonumber\\& \times \operatorname{Tr}\Big(h_{s_I(\Delta)}\left\{h_{s_I(\Delta)}^{-1},\left\{H^E(1), V_{(v(\Delta),\epsilon)}\right\}\right\} \nonumber\\
	&\times h_{s_J(\Delta)}\left\{h_{s_J(\Delta)}^{-1}, V_{(v(\Delta),\epsilon)}\right\} h_{s_K(\Delta)}\left\{h_{s_K(\Delta)}^{-1}, V_{(v(\Delta),\epsilon)}\right\}\Big) \nonumber\\ & \times \sum_{v^{\prime} \! \in \! \beta \!(\! v \!)\!} \frac{1}{E \! \left(\! v^{\prime} \!\right)\!} \sum_{v \! \left(\! \Delta^{\prime} \! \right)\!=v^{\prime}} \epsilon \! \left(\! s_{L^{\prime}} s_{M^{\prime}} s_N^{\prime} \! \right)\! \varepsilon^{L^{\prime} M^{\prime} N^{\prime}} \chi_\epsilon \! \left(\! v\left(\Delta^{\prime}\! \right)\!-\! v \!(\!\Delta \!) \right)\! \nonumber\\& \times \operatorname{Tr}\Big(h_{s_{L^{\prime}}\left(\Delta^{\prime}\right)}\left\{h_{s_{L^{\prime}}\left(\Delta^{\prime}\right)}^{-1}, V_{\left(v\left(\Delta^{\prime}\right),\epsilon \right)}^{\frac{2}{3}}\right\} \nonumber\\ &\times h_{s_{M^{\prime}} \!\left(\! \Delta^{\prime}\! \right)\!} \! \left\{ \! h_{s_{M^{\prime}} \! \left(\! \Delta^{\prime} \!\right)\!}^{-1}, V_{\! \left(\! v\left(\Delta^{\prime}\! \right)\!,\epsilon \!\right)\!}^{\frac{2}{3}}\right\} h_{s_{N^{\prime}} \!\left(\! \Delta^{\prime}\! \right)\!} \!\left\{\! h_{s_{N^{\prime}} \!\left(\! \Delta^{\prime}\! \right)\!}^{-1}, V_{\! \left(\! v\left(\Delta^{\prime}\! \right)\!,\epsilon \!\right)\!}^{\frac{2}{3}}\!\right\}\! \Big) ,
\end{align}
where $E(v)=\left(\begin{array}{c}n(v) \\ 3\end{array}\right)$ is the number of different triples of edges at $v$ and $n(v)$ is the valence of $v$. Since all constituents in this expression have clear quantum analogs, we can define the corresponding operator acting on the basis as
\begin{align}
	\label{C2operator}
	& \widehat{C}_2^{\delta}(N) \cdot|\beta, \vec{j}, \vec{\iota}\rangle \otimes |\phi \rangle \nonumber\\
	&\!=\! i 2^9 \gamma^{-\frac{15}{2}} \! \sum_{v \in \beta(v)} \frac{N(v) D(v)}{E(v)^2}  \!
		\sum_{v(\Delta)=v\left(\Delta^{\prime}\right)=v} \epsilon\left(s_I s_J s_K\right) \! \varepsilon^{I J K} \nonumber\\
	&  \cdot \operatorname{Tr}\left(\hat{h}_{s_I(\Delta)}\left[\hat{h}_{s_I(\Delta)}^{-1},\left[\hat{H}^E(1), \hat{V}_{(v(\Delta),\epsilon)}\right]\right] \right. \nonumber\\
	& \quad \! \cdot \! \left.\hat{h}_{s_J(\Delta)}\left[\hat{h}_{s_J(\Delta)}^{-1}, \hat{V}_{(v(\Delta),\epsilon)}\right] 
		 \hat{h}_{s_K(\Delta)}\left[\hat{h}_{s_K(\Delta)}^{-1}, \hat{V}_{(v(\Delta),\epsilon)}\right]\right) \nonumber\\
	& \cdot \epsilon\left(s_L s_M s_N\right) \varepsilon^{L M N}  \operatorname{Tr}\left(\hat{h}_{s_{L^{\prime}\left(\Delta^{\prime}\right)}}\left[\hat{h}_{s_L\left(\Delta^{\prime}\right)}^{-1}, \hat{V}_{(v(\Delta^{\prime}),\epsilon)}^{\frac{2}{3}}\right] \right. \nonumber\\
	& \quad
	\cdot\left.\hat{h}_{s_{M}\left(\Delta^{\prime}\right)}\left[\hat{h}_{s_M\left(\Delta^{\prime}\right)}^{-1}, \hat{V}_{(v(\Delta^{\prime}),\epsilon)}^{\frac{2}{3}}\right] \right. \nonumber\\
	& \quad \left. \cdot \hat{h}_{s_N\left(\Delta^{\prime}\right)}\left[\hat{h}_{s_N\left(\Delta^{\prime}\right)}^{-1}, \hat{V}_{(v(\Delta^{\prime}),\epsilon)}^{\frac{2}{3}}\right]\right) 
	\cdot |\beta, \vec{j}, \vec{\iota}\rangle \otimes |\phi \rangle.
\end{align}

For the third term $C_3(N)$, we introduce a family of partitions of $\Sigma$, parametrized by $\rho$, such that $\mathrm{sgn}(N(x))$ takes the same value on each cell $\square^{\rho}$, and all cells shrink uniformly as $\rho \rightarrow 0$\cite{domagala2010gravity}. Then, the third term can be expressed as
\begin{align}
\label{C3}
C_3(N) =\mp \lim _{\rho \rightarrow 0}\sum_{\square^{\rho}} \operatorname{sgn}\left(N_{\square^{\rho}}\right) \sqrt{T_{\square^{\rho}}},
\end{align}
where
\begin{align} 
\label{T}
& T_{\square^{\rho}}=-\! \left(\! \int_{\! \square^{\rho}} d^3 x N(x) \sqrt{h}\!\right)\! \!\left(\! \int_{\square^{\rho}}\! d^3 x N(x)(3+2 \omega(\phi)) H^E \!\right)\! \nonumber \\ & +\! \left(\! \int_{\square^{\rho}} d^3 x N \!(\! x \!) \! \sqrt{h}\!\right)\!\left(\!\int_{\square^{\rho}} d^3 x N \!(\! x \!)\!(\!3 \!+\! 2 \omega(\phi)\!)\!\left(\!\gamma^2 \!+\! \frac{1}{\phi^2}\!\right)\! H^L\!\right)\! \nonumber\\ & \!-\! \left(\! \int_{\square^{\rho}} d^3 x N \!(\! x \!)\! \sqrt{h} \!\right)\! \left(\! \int_{\square^{\rho}} d^3 x N \!(\! x \!)\! \frac{(3\!+\! 2 \omega \!(\! \phi \!)\! )\! \omega \!(\! \phi \!)\! }{2 \phi^2} \! \sqrt{h} \!\left(\! \partial_a \phi \!\right)\! \partial^a \phi \!\right)\! \nonumber\\ & -\! \left(\! \int_{\square^{\rho}} d^3 x N(x) \sqrt{h}\! \right)\! \left(\! \int_{\square^{\rho}} d^3 x N(x) \frac{3+2 \omega(\phi)}{\phi} \sqrt{h} D^a D_a \phi \!\right)\! \nonumber\\ & -\left(\int_{\square^{\rho}} d^3 x N(x) \sqrt{\frac{(3+2 \omega(\phi)) \xi(\phi)}{\phi}} \sqrt{h}\right)^2
\end{align}
with
\begin{align}
\label{Lorentz term}
H^L=\frac{1}{2} \frac{\varepsilon_{j m n} \varepsilon_{j k l} \tilde{K}_a^m \tilde{K}_b^n E_k^a E_l^b}{\sqrt{h}}.
\end{align}
Eqation (\ref{T}) is written into five terms as $T_{\square^{\rho}}=\sum_{i=1}^{5} T_{\square^{\rho}}^{i}$. We assume that $\omega(\phi)$ and $\xi(\phi)$ are polynomials in $\phi$ or $\phi^{-1}$, such that their corresponding operators have eigenvalues denoted by $\bar{\omega}$ and $\bar{\xi}$, respectively, on the eigenstates $|\phi\rangle$. Then, the operators corresponding to the terms $T_{\square^{\rho}}^{1},T_{\square^{\rho}}^{2}$ and $T_{\square^{\rho}}^{5}$ act on the basis, respectively, as
\begin{align} 
\label{T1}
&\! \widehat{T_{\square^{\rho}}^{1, \delta}} \! \cdot \! |\beta, \vec{j}, \vec{\iota}\rangle \otimes |\phi \rangle \!=\!-\! 8 \! \sum_{v^{\prime} \in \beta(v) \cap \square^{\rho}} \! \frac{N(v^{\prime})}{E(v^{\prime})} \! \sum_{v(\Delta^{\prime})=v^{\prime}} \! \hat{V}_{(v(\Delta^{\prime}),\epsilon)} \! \nonumber\\
&\cdot \sum_{v \in \beta(v) \cap \square^{\rho}} N(v)(3+2\bar{\omega}(v)) \hat{H}^{E}_{v} \cdot |\beta, \vec{j}, \vec{\iota}\rangle \otimes |\phi \rangle,
\end{align}
\begin{align} 
\label{T2}
&\widehat{T_{\square^{\rho}}^{2, \delta}} \cdot|\beta, \vec{j}, \vec{\iota}\rangle \otimes |\phi \rangle =8\sum_{v^{\prime} \in \beta(v) \cap \square^{\rho}} \frac{N(v^{\prime})}{E(v^{\prime})} \! \sum_{v(\Delta^{\prime})=v^{\prime}}\hat{V}_{(v(\Delta^{\prime}),\epsilon)} \nonumber\\& \cdot \! \sum_{v \in \beta(v) \cap \square^{\rho}} \! N(v)(3 \!+\! 2\bar{\omega}(v)) (\gamma^2 \!+\! D^2(v)) \hat{H}^{L}_{v} \cdot |\beta, \vec{j}, \vec{\iota}\rangle \otimes |\phi \rangle,
\end{align}
\begin{align} 
\label{T5}
&\! \widehat{T_{\square^{\rho}}^{5, \delta}} \! \cdot \! |\beta, \vec{j}, \vec{\iota}\rangle \otimes |\phi \rangle \! =\! -\! 8 \!\sum_{v^{\prime} \in \beta(v) \cap \square^{\rho}} \frac{N(v^{\prime})}{E(v^{\prime})} \sqrt{(3+2\bar{\omega}) \bar{\xi} D(v^{\prime})} \nonumber\\
&\cdot \sum_{v(\Delta^{\prime})=v^{\prime}}\hat{V}_{(v(\Delta^{\prime}),\epsilon)}  \sum_{v \in \beta(v) \cap \square^{\rho}} \frac{N(v)}{E(v)}\sqrt{(3+2\bar{\omega}) \bar{\xi} D(v)} \nonumber\\ &\cdot \sum_{v(\Delta)=v}\hat{V}_{(v(\Delta),\epsilon)} \cdot |\beta, \vec{j}, \vec{\iota}\rangle \otimes |\phi \rangle.
\end{align}
For the term $T_{\square^{\rho}}^{3}$, we use the point-splitting scheme and apply Eqs.~(\ref{E})--~(\ref{K}) to express it as
\begin{align} 
\label{T31}
T_{\square^{\rho}}^{3}=&\left(\int_{\square^\rho} d^3 x N(x) \sqrt{h}\right) \nonumber\\ &\times \int_{\square^{\rho}} d^3 x N(x) \frac{(3+2 \omega) \omega}{2 \phi^2} \frac{1}{\sqrt{h}} E_i^a E_i^b \partial_a \phi \partial_b \phi \nonumber\\ =&\left(\int_{\square^\rho} d^3 x N(x) \sqrt{h}\right) \nonumber\\ &\times \lim _{\epsilon \rightarrow 0} \int_{\square^{\rho}} d^3 x N(x) \frac{(3+2 \omega) \omega}{2 \phi^2} V_{\left(x, \epsilon \right)}^{-\frac{1}{2}} E_i^a \partial_a \phi \nonumber\\
&\times \int d^3 y \chi(x-y) V_{\left(y, \epsilon \right)}^{-\frac{1}{2}} E_i^b \partial_b \phi \nonumber\\  =&\frac{1024}{81} \gamma^{-4} \left(\int_{\square^\rho} d^3 x N(x) \sqrt{h}\right)\nonumber\\
&\times \lim _{\epsilon \rightarrow 0} \int_{\square^{\rho}} d^3 x N(x) \frac{(3+2 \omega) \omega}{2 \phi^2}\left(\partial_a \phi\right) \nonumber\\
&\times \varepsilon^{a c d} \varepsilon_{i j k}\left\{A_c^j(x), V_{(x, \epsilon)}^{\frac{3}{4}}\right\}\left\{A_d^k(x), V_{\left(x,\epsilon \right)}^{\frac{3}{4}}\right\} \nonumber\\ & \times \int d^3 y \chi_\epsilon(x-y) \varepsilon_{i m n} \varepsilon^{b e f}\left(\partial_b \phi\right) \nonumber\\ 
&\times \left\{A_e^m(y), V_{(y,\epsilon)}^{\frac{3}{4}}\right\}\left\{A_f^n(y), V_{\left(y, \epsilon \right)}^{\frac{3}{4}}\right\}.
\end{align}
Then, by Eqs.~(\ref{h}) and ~(\ref{epsilon}), it can be regularized as
\begin{align} 
\label{T32}
&T_{\square^{\rho}}^{3, \delta} =\frac{2^{21}}{3^6}\gamma^{-4} \lim _{\epsilon \rightarrow 0}\sum_{v^{\prime\prime} \in \beta(v) \cap \square^{\rho}} \frac{N(v^{\prime\prime})}{E(v^{\prime\prime})} \sum_{v(\Delta^{\prime\prime})=v^{\prime\prime}} V_{(v(\Delta^{\prime\prime}), \epsilon)} \nonumber \\
&\times \sum_{v \in \beta(v) \cap \square^{\rho}} \frac{N(v)}{E(v)} \frac{(3+2 \omega) \omega}{2 \phi^2}(v) \sum_{v(\Delta)=v} \epsilon\left(s_l s_J s_K\right) \varepsilon^{I J K} \nonumber\\
&\! \times \! \operatorname{Tr} \!\left(\!\tau_i \! h_{s_J(\! \Delta \!)}\!\left\{\! h_{s_J(\! \Delta \!)}^{-1}, V_{(v(\! \Delta \!),\epsilon)}^{\frac{3}{4}}\!\right\}\! h_{s_K(\! \Delta \! )}\!\left\{\! h_{s_K(\Delta)}^{-1}, V_{(v(\! \Delta \!),\epsilon)}^{\frac{3}{4}}\! \right\}\!\right) \! \nonumber\\& \times \Big(\phi\left(f\left(s_I(\Delta)\right)\right)-\phi(v(\Delta))\Big) \nonumber\\ & \! \times \! \sum_{v^{\prime} \in \beta(v)} \! \frac{1}{E\left(v^{\prime}\right)} \! \sum_{v\left(\Delta^{\prime}\right)=v^{\prime}} \! \epsilon\left(s_{L} s_{M} s_N\right) \! \varepsilon^{L M N} \! \chi_\epsilon \! \left(v\left(\Delta^{\prime}\right)\!-\! v(\Delta)\right)\! \nonumber\\& \times \!\operatorname{Tr}\!\left(\!\tau_i \! h_{s_{M^{\prime}}\!\left(\!\Delta^{\prime}\!\right)\!}\!\left\{\! h_{s_M \!\left(\! \Delta^{\prime}\!\right)\!}^{-1}, V_{\!(v \!(\! \Delta^{\prime}\!)\!,\epsilon \!)\!}^{\frac{3}{4}}\!\right\}\! h_{s_N \!\left(\! \Delta^{\prime}\!\right)\!}\!\left\{\! h_{s_N \!\left(\!\Delta^{\prime}\!\right)\!}^{-1}, V_{\!(\! v \!(\!\Delta^{\prime}\!)\!,\epsilon)\!}^{\frac{3}{4}}\! \right\}\!\right)\! \nonumber\\ &\times \Big(\phi(f(s_L(\Delta^{\prime})))-\phi(v(\Delta^{\prime}))\Big), 
\end{align}
where $f(s_{I}(\Delta))$ is the end point of $s_{I}(\Delta)$. Thus, the corresponding operator can be defined as 
\begin{align} 
\label{OT3}
&\widehat{T_{\square^{\rho}}^{3, \delta}} \cdot|\beta, \vec{j}, \vec{\iota}\rangle \otimes |\phi \rangle=\frac{2^{21}}{3^6} \gamma^{-4}\sum_{v^{\prime\prime} \in \beta(v) \cap \square^{\rho}} \frac{N(v^{\prime\prime})}{E(v^{\prime\prime})} \nonumber \\ &\cdot\sum_{v(\Delta^{\prime\prime})=v^{\prime\prime}} \hat{V}_{(v(\Delta^{\prime\prime}), \epsilon)} 
\sum_{v \in \beta(v) \cap \square^{\rho}} \frac{N(v)}{E(v)^2} \frac{(3+2 \bar{\omega}) \bar{\omega} D^2}{2}(v) \nonumber\\&\cdot \! \sum_{v(\Delta)=v(\Delta^{\prime})=v} \epsilon\left(s_I s_J s_K\right) \varepsilon^{I J K}\Big(\phi\left(f\left(s_I(\Delta)\right)\right)\!-\! \phi(v(\Delta))\Big)\nonumber \\&\!\cdot \! \operatorname{Tr}\!\left(\!\tau_i \!\hat{h}_{s_J(\Delta)} \!\left[\!\hat{h}_{s_J(\Delta)}^{-1}, \hat{V}_{(v(\Delta),\epsilon)}^{\frac{3}{4}}\!\right]\! \hat{h}_{s_K(\Delta)}\!\left[\!\hat{h}_{s_K(\Delta)}^{-1}, \hat{V}_{(v(\Delta),\epsilon)}^{\frac{3}{4}}\!\right]\!\right)\! \nonumber \\ & \cdot \epsilon \left(s_L s_M s_N\right) \varepsilon^{L M N}\Big(\phi\left(f\left(s_L\left(\Delta^{\prime}\right)\right)\right)-\phi\left(v\left(\Delta^{\prime}\right)\right)\Big)\nonumber \\& \cdot \operatorname{Tr}\Big(\tau_i \hat{h}_{s_M\left(\Delta^{\prime}\right)}\left[\hat{h}_{s_M\left(\Delta^{\prime}\right)}^{-1}, \hat{V}_{(v(\Delta^{\prime}),\epsilon)}^{\frac{3}{4}}\right]\nonumber\\ &\quad \cdot \hat{h}_{s_N\left(\Delta^{\prime}\right)}\left[\hat{h}_{s_N\left(\Delta^{\prime}\right)}^{-1}, \hat{V}_{(v(\Delta^{\prime}),\epsilon)}^{\frac{3}{4}}\right]\Big) \cdot|\beta, \vec{j}, \vec{\iota}\rangle \otimes |\phi \rangle.
\end{align}
For the term $T_{\square^{\rho}}^{4}$, we add a characteristic function $\chi_{\square^{\rho}}$ of $\square^{\rho}$ to transpose the integral on $\square^{\rho}$ to $\Sigma$. This allows us to use integration by parts and throw away the boundary term. By using the point-splitting scheme and Eqs.~(\ref{E})--~(\ref{K}), this term can be expressed as
\begin{align} 
\label{T41}
T_{\square^{\rho}}^{4}=&\left(\int_{\square^\rho} d^3 x N(x) \sqrt{h}\right)\nonumber \\
&\times \int d^3 x N(x) \chi_{\square^{\rho}}(x) \frac{(3+2 \omega)}{\phi} \sqrt{h} D^a D_a \phi \nonumber \\  =&\left(\int_{\square^\rho} d^3 x N(x) \sqrt{h}\right)\nonumber \\
&\times \lim _{\epsilon \rightarrow 0} \int d^3 x N(x) \chi_{\square^{\rho}}(x) \frac{(3+2 \omega)}{\phi} \sqrt{h} \nonumber\\
&\times \int d^3 y \chi_{\epsilon}(x-y) V_{(y,\epsilon)}^{-1} \sqrt{h} D^a D_a \phi \nonumber\\  =&-\left(\int_{\square^\rho} d^3 x N(x) \sqrt{h}\right) \nonumber\\
&\times \lim _{\epsilon \rightarrow 0} \int d^3 x N(x) \chi_{\square^{\rho}}(x) \frac{(3+2 \omega)}{\phi} \sqrt{h} \nonumber\\
&\times\int d^3 y \partial_a\left(V_{(y,\epsilon)}^{-1} \chi(x-y)\right) \frac{1}{\sqrt{h}} E_i^a E_i^b \partial_b \phi \nonumber\\  =&-\left(\int_{\square^\rho} d^3 x N(x) \sqrt{h}\right) \nonumber\\
&\times\lim _{\epsilon \rightarrow 0} \int d^3 x N(x) \chi_{\square^{\rho}}(x) \frac{(3+2 \omega)}{\phi} \sqrt{h} \nonumber\\
&\!\times \! \int d^3 y \partial_a \!\left(\! V_{(y,\epsilon)}^{-1} \chi_{\epsilon}(x-y)\!\right)\!  \frac{1}{2} \varepsilon^{a c d} \varepsilon_{i m n} e_c^m e_d^n V_{(y,\epsilon)}^{-\frac{1}{2}} \nonumber\\
&\times\int d^3 z \chi_{\epsilon}(y-z)  \frac{1}{2} \varepsilon^{b e f} \varepsilon_{i j k} e_e^j e_f^k V_{(z,\epsilon)}^{-\frac{1}{2}} \partial_b \phi \nonumber\\  =&-\frac{1024}{81} \gamma^{-4}\left(\int_{\square^\rho} d^3 x N(x) \sqrt{h}\right) \nonumber\\
&\times \lim _{\epsilon \rightarrow 0} \int d^3 x N(x) \chi_{\square^{\rho}}(x) \frac{(3+2 \omega)}{\phi} \sqrt{h} \nonumber\\
&\times\int d^3 y \partial_a\left(V_{(y,\epsilon)}^{-1} \chi_{\epsilon}(x-y)\right) \nonumber\\ & \times \varepsilon^{a c d} \varepsilon_{i m n}\left\{A_c^m(x), V_{(y,\epsilon)}^{\frac{3}{4}}\right\}\left\{A_d^n(x), V_{(y,\epsilon)}^{\frac{3}{4}}\right\} \nonumber\\
&\times \int d^3 z \chi_\epsilon(y-z) \varepsilon_{i j k} \varepsilon^{b e f}\left(\partial_b \phi\right)\nonumber\\
&\times \left\{A_e^j(y), V_{(y,\epsilon)}^{\frac{3}{4}}\right\}\left\{A_f^k(y), V_{(y,\epsilon)}^{\frac{3}{4}}\right\}.
\end{align}
Consequently, it can be regularized as
\begin{align} 
\label{T42}
& T_{\square^{\rho}}^{4,\delta}=\!-\!\frac{2^{23}}{3^7}\! \gamma^{-4}\! \lim _{\epsilon \rightarrow 0}\!\sum_{v^{\prime\prime\prime} \in \beta(\! v \!) \cap \square^{\rho}} \frac{N(\! v^{\prime\prime\prime}\!)}{E(\! v^{\prime\prime\prime}\!)} \!\sum_{v(\!\Delta^{\prime\prime\prime}\!)=v^{\prime\prime\prime}} V_{(\! v \!(\! \Delta^{\prime\prime\prime}\!), \epsilon \!)\!}  \nonumber\\
&\times \sum_{v \in \beta(v)} \frac{N(v) \chi_{\square}(v)}{E(v)} \frac{(3+2 \omega)}{\phi}(v) \sum_{v(\Delta)=v} V_{(v(\Delta), \epsilon)} \nonumber\\ & \times \sum_{v^{\prime} \in \beta(v)} \frac{1}{E\left(v^{\prime}\right)} \sum_{v\left(\Delta^{\prime}\right)=v^{\prime}} \epsilon\left(s_L s_M s_N\right) \varepsilon^{L M N}\nonumber\\& \times \Big(V_{\left(f\left(s_L\left(\Delta^{\prime}\right)\right),\epsilon \right)}^{-1} \chi_{\epsilon}\left(v(\Delta)-f\left(s_L\left(\Delta^{\prime}\right)\right)\right) \nonumber\\
&\quad-V_{\left(v\left(\Delta^{\prime}\right),\epsilon \right)}^{-1} \chi_{\epsilon}\left(v(\Delta)-v\left(\Delta^{\prime}\right)\right)\Big) \nonumber\\& \times \operatorname{Tr}\Big(\tau_i h_{s_{M}\left(\Delta^{\prime}\right)}\left\{h_{s_{M}\left(\Delta^{\prime}\right)}^{-1}, V_{\left(v\left(\Delta^{\prime}\right),\epsilon \right)}^{\frac{3}{4}}\right\} \nonumber\\
&\quad \times h_{s_N\left(\Delta^{\prime}\right)}\left\{h_{s_N\left(\Delta^{\prime}\right)}^{-1}, V_{\left(v\left(\Delta^{\prime}\right),\epsilon \right)}^{\frac{3}{4}}\right\}\Big) \nonumber\\ & \times \sum_{v^{\prime \prime} \in \beta(v)} \frac{1}{E\left(v^{\prime \prime}\right)} \sum_{v\left(\Delta^{\prime \prime}\right)=v^{\prime \prime}}  \chi_\epsilon\left(v\left(\Delta^{\prime}\right)-v\left(\Delta^{\prime \prime}\right)\right) \nonumber\\& \times \epsilon\left(s_I s_J s_K\right) \varepsilon^{I J K}\operatorname{Tr}\Big(\tau_{i} h_{s_J\left(\Delta^{\prime \prime}\right)}\left\{h_{s_J\left(\Delta^{\prime \prime}\right)}^{-1}, V_{\left(v\left(\Delta^{\prime \prime}\right), \epsilon\right)}^{\frac{3}{4}}\right\}\nonumber\\
&\quad\times h_{s_K\left(\Delta^{\prime \prime}\right)}\left\{h_{s_K\left(\Delta^{\prime \prime}\right)}^{-1} V_{\left(v\left(\Delta^{\prime \prime}\right), \epsilon\right)}^{\frac{3}{4}}\right\}\Big) \nonumber\\
&\times \Big(\phi\left(f\left(s_I\left(\Delta^{\prime \prime}\right)\right)\right)-\phi\left(v\left(\Delta^{\prime \prime}\right)\right)\Big). 
\end{align}
Thus, the corresponding operator can be defined as
\begin{align} 
\label{OT4}
&\widehat{T_{\square^{\rho}}^{4, \delta}} \cdot|\beta, \vec{j}, \vec{\iota}\rangle \otimes |\phi \rangle=\frac{2^{23}}{3^7} \gamma^{-4} \sum_{v^{\prime\prime\prime} \in \beta(v) \cap \square^{\rho}} \frac{N(v^{\prime\prime\prime})}{E(v^{\prime\prime\prime})} \nonumber \\ &\cdot \sum_{v(\Delta^{\prime\prime\prime})=v^{\prime\prime\prime}} \hat{V}_{(v(\Delta^{\prime\prime\prime}), \epsilon)} \sum_{v \in \beta(v)} \frac{N(v) \chi_{\square}(v)(3+2 \bar{\omega} )D(v)}{E(v)^3}  \nonumber\\&\cdot \sum_{v(\Delta)=v\left(\Delta^{\prime}\right)=v\left(\Delta^{\prime\prime}\right)=v} \epsilon\left(s_{L} s_{M} s_N\right) \varepsilon^{L M N} \nonumber\\
&\cdot \operatorname{Tr}\Big(\tau_i \hat{h}_{s_M\left(\Delta^{\prime}\right)}\left[\hat{h}_{s_M\left(\Delta^{\prime}\right)}^{-1}, \hat{V}_{\left(v\left(\Delta^{\prime}\right), \epsilon\right)}^{\frac{3}{4}}\right] \nonumber\\ &\quad \cdot \hat{h}_{s_N\left(\Delta^{\prime}\right)}\left[\hat{h}_{s_N\left(\Delta^{\prime}\right)}^{-1}, \hat{V}_{\left(v\left(\Delta^{\prime}\right), \varepsilon\right)}^{\frac{3}{4}}\right]\Big)\nonumber \\ & \cdot \epsilon\left(s_I s_J s_K\right) \varepsilon^{I J K}\Big(\phi\left(f\left(s_I\left(\Delta^{\prime \prime}\right)\right)\right)-\phi\left(v\left(\Delta^{\prime \prime}\right)\right)\Big)\nonumber \\&\cdot \operatorname{Tr}\Big(\tau_i \hat{h}_{s_J\left(\Delta^{\prime \prime}\right)}\left[\hat{h}_{s_J\left(\Delta^{\prime \prime}\right)}^{-1}, \hat{V}_{\left(v\left(\Delta^{\prime \prime}\right) ,\epsilon \right)}^{\frac{3}{4}}\right]\nonumber\\ &\quad \cdot \hat{h}_{s_K\left(\Delta^{\prime \prime}\right)}\left[\hat{h}_{s_K\left(\Delta^{\prime \prime}\right)}^{-1}, \hat{V}_{\left(v\left(\Delta^{\prime \prime}\right), \epsilon\right)}^{\frac{3}{4}}\right]\Big) \cdot|\beta, \vec{j}, \vec{\iota}\rangle \otimes |\phi \rangle.
\end{align}
Since $\widehat{T_{\square^{\rho}}}$ acts on the vertices, the operator $\hat{C}_{3}^{\delta}(N)$ can be expressed as a sum over vertices when $\rho \rightarrow 0$ as
\begin{align}
\label{OC4}
\! \hat{C}_3^{\delta}(\! N \!) \! \cdot \! |\beta,\!  \vec{j}, \! \vec{\iota}\rangle \! \otimes \! |\phi \rangle \!=\! \mp \! \sum_{v} sgn(\! N_v \!) \sqrt{\! \sum_{i=1}^{5}\widehat{T^{i, \delta}_{\square^{\rho,v}}}} \! \cdot \! |\beta, \! \vec{j}, \! \vec{\iota}\rangle \! \otimes \! |\phi \rangle \!
\end{align}
where $\square^{\rho,v}$ is the cell contain the vertex $v$.

We have obtained the regulated operator $\hat{C}^{\delta}(N)\equiv \hat{C}_{1}(N)+\hat{C}_{2}^{\delta}(N)+\hat{C}_{3}^{\delta}(N)$ in the kinematic Hilbert space $\mathcal{H}_{\mathrm{kin} }^{t}$. To remove the regulator $\delta$, we employ it on the vertex Hilbert space introduced in Refs.\cite{yang2015new,lewandowski2015symmetric,li2025loop}. Note that the operators $\hat{C}_2^{\delta}(N)$ and $\hat{C}_3^{\delta}(N)$ act only on the noncoplanar vertices with valence greater than 3 for a given cylindrical function over graph $\beta$. The set of these vertices is denoted by $V_{np4}(\beta)$. One defines the antilinear rigging map $\tilde{\eta}:Cyl_{G} \rightarrow Cyl^{\star}$ as 
\begin{align}
	\label{rigging map}
	\tilde{\eta}\left(\Psi_\beta \right)\left[\Phi_{\beta^{\prime}}\right]:=\sum_{g \in Diff_{np4} / TD i f f_\beta}\left\langle\hat{U}_{g} \Psi_\beta \mid \Phi_{\beta^{\prime}}\right\rangle_{kin},
\end{align}
where $Diff_{np4}$ is the group of all diffeomorphisms preserving each vertex of $V_{np4}(\beta)$ while $TD i f f_\beta$ is the subgroup of $Diff_{np4}$ consisting of those elements that preserve each edge of $\beta$. Then, the inner product on $\tilde{\eta}(Cyl_{G})$ can be defined by
\begin{align}
	\label{scalar product5}
	\left<\tilde{\eta}(\Psi_\beta)|\tilde{\eta}(\Phi_{\beta^{\prime}})\right>_{vtx}:=\tilde{\eta}\left(\Psi_\beta \right)\left[\Phi_{\beta^{\prime}}\right].
\end{align}
By completing $\tilde{\eta}(Cyl_{G})$ using Eq.~(\ref{scalar product5}), one obtains the vertex Hilbert space $\mathcal{H}_{vtx}$. It turns out that $\hat{C}^{\delta}(N)$ can be promoted to an $\delta$-independent operator on $\mathcal{H}_{vtx}$ via the following dual action:
\begin{align}
	[\hat{C}^{*}(N)\cdot \tilde{\eta}(\Psi_\beta)]\cdot \Phi_{\beta^{\prime}}:=\tilde{\eta}(\Psi_\beta)(\hat{C}^{\delta}(N)\cdot \Phi_{\beta^{\prime}}).
\end{align}

We introduce the notation $(\Psi |\alpha,\phi\rangle\equiv \tilde{\eta}(\Psi)[|\alpha,\phi\rangle]$ and then impose the constraint equation 
\begin{align}
	\label{cp}
	( \hat{C}^{*}(N)\cdot\Psi  |\alpha,\phi\rangle=0 \ \ \ \ \ \ \ \forall \ \ |\alpha,\phi \rangle,
\end{align} 
where we denote $|\beta, \vec{j}, \vec{\iota}\rangle \otimes|\phi\rangle \equiv |\alpha,\phi\rangle$. By defining $\hat{H_p}(N):=-\hat{C_2^\delta}(N)-\hat{C_3^\delta}(N)$, Eq.~(\ref{cp}) can be written as
\begin{align}
	\label{ch}
	( \Psi | \hat{C_1}(N)|\alpha,\phi\rangle=( \Psi | \hat{H_p}(N)|\alpha,\phi\rangle.
\end{align} 
Substituting Eq.(\ref{term1}) into Eq.~(\ref{ch}), we have
\begin{align}
	\frac{1}{2i \lambda_0} \! \Big(\! (\Psi|\alpha,\phi \!-\! \lambda_0 N \rangle \!-\! (\Psi|\alpha,\phi \!+\! \lambda_0 N \rangle \!\Big)\!=\!( \! \Psi | \hat{H_p}(N)|\alpha,\phi\rangle \! \label{c12}.
\end{align} 
It should be noted that Eq.~(\ref{c12}), which arises when performing polymer quantization with respect to $\pi$, can be regarded as the Schrödinger-like difference equation. It describes the discrete time evolution of physical states with respect to $\phi$, where $\hat{H_p}(N)$ plays the role of the physical Hamiltonian operator. Although it is challenging to provide a general solution of Eq.~(\ref{c12}), it is possible to obtain specific solutions when applying this deparametrized framework of scalar-tensor theory to certain symmetry-reduced models. In the next section, we will apply this framework to a cosmological model and construct explicit dynamics based on the solution to the deparametrized Hamiltonian constraint equation.

In the sector of $\omega(\phi) =-\frac{3}{2}$, the conformal constraint can be quantized as 
\begin{align} 
\label{OC}
&\hat{S}(\lambda) \cdot |\beta, \vec{j}, \vec{\iota}\rangle \otimes |\phi \rangle \nonumber \\=&\frac{-1}{2 i \lambda_0}\Big(\exp \left[i \lambda_0 \int d^3 x \lambda(x) \hat{\pi}(x)\hat{\phi}(x)\right]\nonumber \\
&\quad -\exp \left[-i \lambda_0 \int d^3 x \lambda(x) \hat{\pi}(x)\hat{\phi}(x)\right]\Big) \cdot |\beta, \vec{j}, \vec{\iota}\rangle \otimes |\phi \rangle \nonumber\\&-\int d^3 x \lambda(x) \hat{\tilde{K}}^{i}_{a}\hat{E}^a_i \cdot |\beta, \vec{j}, \vec{\iota}\rangle \otimes |\phi \rangle \nonumber\\=&\frac{1}{2 i \lambda_0}\left(|\beta, \vec{j}, \vec{\iota}\rangle \otimes |\phi +\lambda_0 \lambda \phi \rangle - |\beta, \vec{j}, \vec{\iota}\rangle \otimes |\phi -\lambda_0 \lambda \phi \rangle \right) \nonumber
\\&-i 2^9 \gamma^{-\frac{15}{2}} \sum_{v \in \beta(v)} \frac{\lambda(v)}{E(v)^2} \sum_{v(\Delta)=v\left(\Delta^{\prime}\right)=v} \epsilon\left(s_I s_J s_K\right) \varepsilon^{I J K} \nonumber\\ &\! \cdot \! \operatorname{Tr}\!\Big(\! \hat{h}_{s_I(\Delta)}\left[\hat{h}_{s_I(\Delta)}^{-1},\left[\hat{H}^E(1), \hat{V}_{(v(\Delta),\epsilon)}\right]\right] \nonumber\\ &\quad \cdot  \hat{h}_{s_J(\Delta)}\left[\hat{h}_{s_J(\Delta)}^{-1}, \hat{V}_{(v(\Delta),\epsilon)}\right] \hat{h}_{s_K(\Delta)}\left[\hat{h}_{s_K(\Delta)}^{-1}, \hat{V}_{(v(\Delta),\epsilon)}\right]\!\Big) \! \nonumber\\& \cdot \epsilon\left(s_L s_M s_N\right) \varepsilon^{L M N}  \operatorname{Tr}\Big(\hat{h}_{s_{L^{\prime}\left(\Delta^{\prime}\right)}}\left[\hat{h}_{s_L\left(\Delta^{\prime}\right)}^{-1}, \hat{V}_{(v(\Delta^{\prime}),\epsilon)}^{\frac{2}{3}}\right]  \nonumber \\
&\quad \cdot\hat{h}_{s_{M}\left(\Delta^{\prime}\right)}\left[\hat{h}_{s_M\left(\Delta^{\prime}\right)}^{-1}, \hat{V}_{(v(\Delta^{\prime}),\epsilon)}^{\frac{2}{3}}\right] \nonumber\\ &\quad \cdot\hat{h}_{s_N\left(\Delta^{\prime}\right)}\left[\hat{h}_{s_N\left(\Delta^{\prime}\right)}^{-1}, \hat{V}_{(v(\Delta^{\prime}),\epsilon)}^{\frac{2}{3}}\right]\Big) \cdot|\beta, \vec{j}, \vec{\iota}\rangle \otimes |\phi \rangle .
\end{align}
To quantize the deparametrized Hamiltonian constraint (\ref{new constraint2}), we divide it into the following two terms:
\begin{equation}
\begin{aligned}
\label{two terms}
&C_1^{\prime}(N)=\int d^3 x N(x) \pi(x), \\& C_2^{\prime}(N)=\mp \int d^3 x N(x) \sqrt{t^{\prime}(\phi,A,E)}.
\end{aligned}
\end{equation}
Note that the first term $C_1^{\prime}(N)$ has been quantized as (\ref{term1}). The second term $C_2^{\prime}(N)$ can be expressed as
\begin{equation}
\label{two terms}
C_2^{\prime}(N)=\mp \lim _{\rho \rightarrow 0} \sum_{\square^\rho} \operatorname{sgn}\left(N_{\square^\rho}\right) \sqrt{T^{\prime}_{\square^\rho}},
\end{equation}
where 
\begin{align} 
\label{Tprime}
& T^{\prime}_{\square^{\rho}}=\left(\int_{\square^{\rho}} d^3 x N(x) \sqrt{h}\right)\left(\int_{\square^{\rho}} d^3 x 2 N(x) H^E\right) \nonumber\\ & \!+\!  \left(\!\int_{\square^{\rho}} \! d^3 x N(x)\! (\!\gamma^2 \!+\! \phi^{-2} \!)\! \delta_{u}^l \tilde{K}^u_a E^a_k \!\right)\! \left(\! \int_{\square^{\rho}} \! d^3 x N \!(\! x \!)\! \delta_{v}^k \tilde{K}^v_b E^b_l \!\right)\! \nonumber \\ & \!+\!  \left(\! \int_{\square^{\rho}} d^3 x N(x) (\phi^{-2}-\gamma^2) \tilde{K}^k_a E^a_k \!\right)\! \left(\! \int_{\square^{\rho}} d^3 x N(x) \tilde{K}^l_b E^b_l \!\right) \nonumber
\\ & -\left(\int_{\square^{\rho}} d^3 x N(x) \sqrt{h}\right)\left(\int_{\square^{\rho}} d^3 x N(x) \frac{3}{4 \phi^2} \sqrt{h}\left(\partial_a \phi\right) \partial^a \phi\right) \nonumber \\ & +\left(\int_{\square^{\rho}} d^3 x N(x) \sqrt{h}\right)\left(\int_{\square^{\rho}} d^3 x N(x) \frac{1}{\phi} \sqrt{h} D^a D_a \phi\right) \nonumber \\ & +\left(\int_{\square^{\rho}} d^3 x N(x) \sqrt{\frac{\xi(\phi)}{\phi}} \sqrt{h}\right)^2.
\end{align}
Eqation (\ref{Tprime}) is written into six terms as $T^{\prime}_{\square^{\rho}}=\sum_{i=1}^{6} T^{\prime,i}_{\square^{\rho}}$. Notice that $T^{\prime,1}_{\square^{\rho}},T^{\prime,4}_{\square^{\rho}},T^{\prime,5}_{\square^{\rho}}$ and $T^{\prime,6}_{\square^{\rho}}$ take the same forms as the term $T^{1}_{\square^{\rho}},T^{3}_{\square^{\rho}},T^{4}_{\square^{\rho}}$ and $T^{5}_{\square^{\rho}}$, respectively, in the sector
of $\omega(\phi) \neq-\frac{3}{2}$ and, hence, have been quantized. To quantize the term $T_{\square^{\rho}}^{\prime,2}$, we use Eqs.~(\ref{E})--~(\ref{K}) to express it as
\begin{align} 
\label{Tprime2}
&T_{\square^{\rho}}^{\prime,2}=4 \gamma^{-9}\Big(\int_{\square^{\rho}} d^3 x\left(\gamma^2 \!+\! \phi^{-2}\right) N \delta_{u}^l \left\{A_a^u,\left\{H^E(1), V\right\}\right\} \nonumber\\
&\quad \times \varepsilon_{k m n} \varepsilon^{a c d}\left\{A_c^m, V\right\}\left\{A_d^n, V\right\}\Big) \nonumber\\& \! \times \! \left(\! \int \! d^3 x  N \! \delta_{v}^k \! \left\{\! A_b^v,\left\{H^E(1), V\right\} \!\right\}\! \! \varepsilon_{l i j} \! \varepsilon^{b e f}\left\{A_e^i, V\right\}\{A_f^j, V\}\!\right)\!.
\end{align}
Then it can be regularized by using Eqs.~(\ref{h}) and ~(\ref{epsilon}) as
\begin{align} 
\label{Tprime22}
T_{\square^{\rho}}^{\prime,2, \delta}=&\sum_{v \in \beta(v) \cap \square^{\rho}} \frac{2^{10}}{9} \gamma^{-9} \frac{N(v)}{E(v)}\left(\gamma^2+\phi^{-2}\right)(v) \nonumber\\ &\times \sum_{v(\Delta)=v} \epsilon\left(s_I s_J s_K\right) \varepsilon^{I J K}\nonumber \\&\times\operatorname{Tr}\left(\tau_l h_{s_I(\Delta)}\left\{h_{s_I(\Delta)}^{-1},\left\{H^E(1), V_{(v(\Delta),\epsilon)}\right\}\right\}\right) \nonumber\\
&\times\operatorname{Tr}\Big(\tau_k h_{s_J(\Delta)}\left\{h_{s_J(\Delta)}^{-1}, V_{(v(\Delta),\epsilon)}\right\} \nonumber\\
&\quad \times h_{s_K(\Delta)}\left\{h_{s_K(\Delta)}^{-1}, V_{(v(\Delta),\epsilon)}\right\}\Big) \nonumber\\& \times \sum_{v^{\prime} \in \beta(v) \cap \square^{\rho}} \frac{N\left(v^{\prime}\right)}{E\left(v^{\prime}\right)} \sum_{v\left(\Delta^{\prime}\right)=v^{\prime}} \epsilon\left(s_{L} s_M s_N\right) \varepsilon^{L M N} \nonumber\\
&\times \operatorname{Tr}\left(\tau_k h_{s_{L}(\Delta^{\prime})}\left\{h_{s_{L}(\Delta^{\prime})}^{-1},\left\{H^E(1), V_{(v(\Delta^{\prime}),\epsilon)}\right\}\right\}\right) \nonumber\\& \times\operatorname{Tr}\Big(\tau_l h_{s_M\left(\Delta^{\prime}\right)}\left\{h_{s_{M}\left(\Delta^{\prime}\right)}^{-1}, V_{(v(\Delta^{\prime}),\epsilon)}\right\} \nonumber\\
&\quad \times h_{s_N\left(\Delta^{\prime}\right)}\left\{h_{s_N\left(\Delta^{\prime}\right)}^{-1}, V_{(v(\Delta^{\prime}),\epsilon)}\right\}\Big).
\end{align}
Thus, the corresponding operator can be defined as 
\begin{align} 
\label{OTprime2}
&\!\widehat{T_{\square^{\rho}}^{\prime,2, \delta}}\! \cdot \! |\beta, \vec{j}, \vec{\iota}\rangle \otimes |\phi \rangle \!=\! \sum_{v \in \beta(v) \cap \square^{\rho}}\! \frac{2^{10}}{9} \gamma^{-9} \! \frac{N(v)}{E(v)}\left(\! \gamma^2 \!+\! D^2 \! \right)\!(\! v \!)\! \nonumber \\
& \cdot \sum_{v(\Delta)=v} \epsilon\left(s_I s_J s_K\right) \varepsilon^{I J K} \nonumber \\&\cdot\operatorname{Tr}\left(\tau_l \hat{h}_{s_I(\Delta)}\left[\hat{h}_{s_I(\Delta)}^{-1},\left[\hat{H}^E(1), \hat{V}_{(v(\Delta),\epsilon)}\right]\right]\right) \nonumber\\
&\cdot \operatorname{Tr}\Big(\tau_k \hat{h}_{s_J(\Delta)}\left[\hat{h}_{s_J(\Delta)}^{-1}, \hat{V}_{(v(\Delta),\epsilon)}\right] \nonumber\\
&\quad \cdot \hat{h}_{s_K(\Delta)}\left[\hat{h}_{s_K(\Delta)}^{-1}, \hat{V}_{(v(\Delta),\epsilon)}\right]\Big) \nonumber\\& \cdot \sum_{v^{\prime} \in \beta(v) \cap \square^{\rho}} \frac{N\left(v^{\prime}\right)}{E\left(v^{\prime}\right)} \sum_{v\left(\Delta^{\prime}\right)=v^{\prime}} \epsilon\left(s_{L} s_M s_N\right) \varepsilon^{L M N} \nonumber\\
&\cdot \operatorname{Tr}\left(\tau_k \hat{h}_{s_{L}(\Delta^{\prime})}\left[\hat{h}_{s_{L}(\Delta^{\prime})}^{-1},\left[\hat{H}^E(1), \hat{V}_{(v(\Delta^{\prime}),\epsilon)}\right]\right]\right) \nonumber\\& \cdot \operatorname{Tr}\Big(\tau_l \hat{h}_{s_M\left(\Delta^{\prime}\right)}\left[\hat{h}_{s_{M}\left(\Delta^{\prime}\right)}^{-1}, \hat{V}_{(v(\Delta^{\prime}),\epsilon)}\right] \nonumber\\
&\quad \cdot\hat{h}_{s_N\left(\Delta^{\prime}\right)}\left[\hat{h}_{s_N\left(\Delta^{\prime}\right)}^{-1}, \hat{V}_{(v(\Delta^{\prime}),\epsilon)}\right]\Big) \cdot|\beta, \vec{j}, \vec{\iota}\rangle \otimes |\phi \rangle.
\end{align}
The same method can be applied to quantize the term $T_{\square^{\rho}}^{\prime,3}$ as
\begin{align} 
\label{OTprime3}
&\widehat{T_{\square^{\rho}}^{\prime,3, \delta}} \cdot |\beta, \vec{j}, \vec{\iota}\rangle \otimes |\phi \rangle \!=\! \sum_{v \in \beta(v) \cap \square^{\rho}} \! \frac{2^{10}}{9} \gamma^{-9} \! \frac{N(v)}{E(v)}\! \left( \! D^2 \!-\! \gamma^2\! \right)\!(\! v \!)\!  \nonumber\\
&\cdot \sum_{v(\Delta)=v} \epsilon\left(s_I s_J s_K\right) \varepsilon^{I J K} \nonumber\\&\cdot \operatorname{Tr}\Big(\hat{h}_{s_I(\Delta)}\left[\hat{h}_{s_I(\Delta)}^{-1},\left[\hat{H}^E(1), \hat{V}_{(v(\Delta),\epsilon)}\right]\right] \nonumber\\
&\quad \cdot \hat{h}_{s_J(\Delta)}\left[\hat{h}_{s_J(\Delta)}^{-1}, \hat{V}_{(v(\Delta),\epsilon)}\right] \hat{h}_{s_K(\Delta)}\left[\hat{h}_{s_K(\Delta)}^{-1}, \hat{V}_{(v(\Delta),\epsilon)}\right]\Big) \nonumber\\& \cdot \sum_{v^{\prime} \in \beta(v) \cap \square^{\rho}} \frac{N\left(v^{\prime}\right)}{E\left(v^{\prime}\right)} \sum_{v\left(\Delta^{\prime}\right)=v^{\prime}} \epsilon\left(s_{L} s_M s_N\right) \varepsilon^{L M N} \nonumber\\& \cdot \operatorname{Tr}\Big(\hat{h}_{s_{L}(\Delta^{\prime})}\left[\hat{h}_{s_{L}(\Delta^{\prime})}^{-1},\left[\hat{H}^E(1), \hat{V}_{(v(\Delta^{\prime}),\epsilon)}\right]\right] \nonumber\\
&\quad \cdot \hat{h}_{s_M\left(\Delta^{\prime}\right)}\left[\hat{h}_{s_{M}\left(\Delta^{\prime}\right)}^{-1}, \hat{V}_{(v(\Delta^{\prime}),\epsilon)}\right] \nonumber\\ &\quad \cdot \hat{h}_{s_N\left(\Delta^{\prime}\right)}\left[\hat{h}_{s_N\left(\Delta^{\prime}\right)}^{-1}, \hat{V}_{(v(\Delta^{\prime}),\epsilon)}\right]\Big) \cdot |\beta, \vec{j}, \vec{\iota}\rangle \otimes |\phi \rangle.
\end{align}
Since the operator $\widehat{T^{\prime, \delta}_{\square^{\rho}}}$ acts on the vertices, the resulting operator $\hat{C}_{2}^{\prime}(N)$ can be expressed as a sum over vertices when $\rho \rightarrow 0$ as
\begin{align}
\label{OC4p}
\! \hat{C}_2^{' \delta} \!(\! N \!) \! \cdot \! |\beta,\! \vec{j},\! \vec{\iota}\rangle \! \otimes \! |\phi \rangle \!=\! \mp \! \sum_{v} \! sgn \!(\! N_v \!) \! \sqrt{\sum_{i=1}^{6}\widehat{T^{\prime,i, \delta}_{\square^{\rho,v}}}} \! \cdot \! |\beta, \! \vec{j}, \! \vec{\iota}\rangle \! \otimes \! \cdot|\phi \rangle \!.
\end{align}

Just as in the sector of $\omega(\phi) \neq-\frac{3}{2}$, the operator $\hat{C}^{'\delta}(N)\equiv \hat{C}_{1}^{'}(N)+\hat{C}_{2}^{'\delta}(N)$ can be promoted to an $\delta$-independent operator on the vertex Hilbert space $\mathcal{H}_{vtx}$. Then, the Hamiltonian constraint equation can be written as
\begin{align}
	\label{cp2}
	( \hat{C}^{'*}(N) \cdot \Psi  |\alpha,\phi\rangle=0  \ \ \ \ \ \ \ \forall \ \ |\alpha,\phi \rangle.
\end{align} 
By defining $\hat{H_p^{'}}(N):=-\hat{C_2^{' \delta}}(N)$, we have
\begin{align}
	\frac{1}{2i \lambda_0} \! \Big( \! (\Psi|\alpha,\phi \!-\! \lambda_0 N \rangle \!-\! (\Psi|\alpha,\phi \!+\! \lambda_0 N \rangle \!\Big)\!=\! ( \Psi | \hat{H_p^{'}}(N)|\alpha,\phi\rangle \label{c122}.
\end{align} 
Again, this Schrödinger-like difference equation describe the discrete time evolution of physical states with respect to $\phi$, where $\hat{H_p^{'}}(N)$ plays the role of the physical Hamiltonian operator.

\section{Loop quantum dynamics of Brans-Dicke cosmology}
\label{alqbdc}

The simplified models of quantum gravity serve as practical tools to test the ideas and structures of the full theory and to extract physical predictions. The original scalar-tensor theory of gravity is the so-called Brans-Dicke theory, in which the coupling parameter $\omega$ is a constant and the potential $\xi$ vanishes. The cosmological models of classical Brans-Dicke theory were first studied in Ref.\cite{greenstein1968brans}. Note that the coupling constant $\omega$ of the Brans-Dicke theory is constrained to be a relatively big value by the Solar System experiments\cite{will2014confrontation,will2018theory}. Thus, we will focus on the cosmological model of the Brans-Dicke theory with the coupling constant $\omega \neq -\frac{3}{2}$. 

\subsection{Classical Brans-Dicke cosmology}
We consider the spatially flat, homogeneous and isotropic
cosmological model, where the spacetime metric is described by the so-called Friedman-Robertson-Walker metric:
\begin{align}
\label{FRW}
d s^2=-d t^2+a^2(t)\left(d r^2+r^2\left(d \theta^2+\sin ^2 \theta d \varphi^2\right)\right),
\end{align}
where $a$ is the scale factor. In the cosmological model, the integration over the entire space results in divergence. To overcome this issue, one introduces an elemental cell $V$ in the spatial manifold $\mathbb{R}^3$ and constrains the integrals to this cell. We choose a fiducial Euclidean metric ${^{o}h_{ab}}$ such that the physical spatial metric can be written as $h_{ab}= a^{2} \  {^{o}h_{ab}}$, along with a fiducial triad ${^{o}e^{a}_{i}}$ and the fiducial cotriad ${^{o}\omega_{a}^{i}}$. For simplicity, we assume that the elemental cell $V$ is cubic, as measured by the fiducial metric, and has a volume $V_0$. We also let ${^{o}e^{a}_{i}}$ be the tangent vector of the ith edge of the elemental cell $V$. By fixing the degrees of freedom of local gauge and diffeomorphism transformations, one finally obtains the reduced canonical variables as\cite{ashtekar2003mathematical}
\begin{align}
\label{reduced canonical variables}
A_a^i=c V_o^{-\frac{1}{3} } {^{o}\omega_a^i} &, \ E_i^a=p V_o^{-\frac{2}{3}}{\sqrt{\operatorname{det}\left({^o h_{a b}}\right)}}{^{o} e_i^a}, \nonumber
\\ \phi(x)=\phi &, \ \pi(x)=\tilde{\pi} V_o^{-1},
\end{align}
where $c,p,\phi$, and $\tilde{\pi}$ are constants on the spatial manifold and the new variables are related to the original ones by $c=\left(\gamma \phi \dot{a}+\gamma \frac{a}{2} \dot{\phi}\right) V_0^{\frac{1}{3}}$ and $|p|=a^2 V_0^{\frac{2}{3}}$. Thus, the phase space of the cosmological model consists of conjugate pairs $(c,p)$ and $(\phi,\tilde{\pi})$. The nonzero Poisson brackets between them are given by
\begin{align}
\label{reduced Poisson bracket}
\{c, p\}=\frac{\gamma}{3},\  \{\phi, \tilde{\pi}\}=1.
\end{align}
Since the Gaussian and diffeomorphism constraints have been satisfied in the cosmological model, we need only to address the remaining reduced Hamiltonian constraint. Because of the spatial flatness, homogeneity, and isotropy, all spatial derivatives vanish, and, hence, one has
\begin{align}
\label{reduced variables}
\Gamma_a^i=0, \ \partial_b A_a^i=0, \ A_a^i=\gamma \tilde{K}_a^i,  \ F_{a b}^i=\varepsilon_{j k}^i A_a^j A_b^k.
\end{align}
Then, the Hamiltonian constraint in Eq.~(\ref{Hamiltonianconstraint2}) is reduced to 
\begin{align}
\label{reduced Hamiltonian constraint}
H=-\frac{3 c^2 \sqrt{|p|}}{\gamma^2  \phi}+\frac{1}{(3+2 \omega) \phi|p|^{\frac{3}{2}}}\left(\frac{3 c p}{\gamma}+\tilde{\pi} \phi\right)^2.
\end{align}

The evolution equation of the scalar field reads 
\begin{align}
\label{reduced evolution equation of the scalar field}
\dot{\phi}=\{\phi, H\}=\frac{2}{(3+2 \omega)|p|^{\frac{3}{2}}}\left(\frac{3 c p}{\gamma}+\pi \phi\right) .
\end{align}
A straightforward calculation gives
\begin{align}
\label{evolution constant}
\left\{\frac{3 c p}{\gamma}+\pi \phi, H\right\} & =-\frac{1}{2}\Big(-\frac{6 \omega c^2 \sqrt{|p|}}{(3+2 \omega) \gamma^2 \phi}+\frac{6 c \pi \ \mathrm{sgn}(p)}{(3+2 \omega) \gamma \sqrt{|p|}} \nonumber\\
&\quad+\frac{\pi^2 \phi}{(3+2 \omega)|p|^{\frac{3}{2}}}\Big) \nonumber\\
& =-\frac{1}{2} H \approx 0,
\end{align}
where $\mathrm{sgn}(p)$ is the sign function of $p$. This indicates that the scalar field $\phi$ is a monotonic function with respect to the cosmological time\cite{zhang2012loop}. Therefore, the scalar field can be interpreted as an emergent internal time variable, which can be employed to deparametrize the Hamiltonian constraint.

\subsection{Structure of the quantum theory}
The kinematical Hilbert space of the geometric part can be constructed as $\mathcal{H}_{\mathrm{kin}}^{\mathrm{geo}}\equiv L^2\left(R_{B o h r}, d \mu_H\right)$, which is the space of square-integrable functions on the Bohr compactification of the real line\cite{ashtekar2003mathematical}. The eigenstates of the operator $\hat{p}$ corresponding to the momentum are labeled by real numbers $\mu$ as 
\begin{align}
\label{Eigenstates of p}
\hat{p} \cdot|\mu\rangle=\frac{\gamma}{6} \mu|\mu\rangle.
\end{align}
These states contribute an orthonormal basis in $\mathcal{H}_{\mathrm{kin}}^{\mathrm{geo}}$ by
\begin{align}
\label{basis1}
\left\langle\mu_1 \mid \mu_2\right\rangle=\delta_{\mu_1, \mu_2},
\end{align}
where $\delta_{\mu_1, \mu_2}$ is the Kronecker delta function. Hence, a typical state in $\mathcal{H}_{\mathrm{kin}}^{\mathrm{geo}}$ can be expressed as a countable sum $|\Psi\rangle=\sum_n c_n\left|\mu_n\right\rangle$. For the convenience in studying quantum dynamics using the $\bar{\mu}$ scheme\cite{ashtekar2006quantum}, we define the following new variables:
\begin{align}
\label{newbasis}
b:=\bar{\mu} c,\ v:=2 \sqrt{3} \ \mathrm{sgn}(p) \ \bar{\mu}^{-3},
\end{align}
where $\bar{\mu}=\sqrt{\frac{\Delta}{|p|}}$, with $\Delta=4 \sqrt{3} \pi \gamma G \hbar$ being the minimum nonzero eigenvalue of the area operator. The value of $\bar{\mu}$ represents the ratio of the side length of the minimum area to the side length of the elemental cell. These variables also form a conjugate pair, and their nonzero Poisson bracket reads 
\begin{align}
\label{b variables bv}
\{b, v\}=2.
\end{align}
It turns out that the eigenstates $|v\rangle$ of $\hat{v}$ also contribute an orthonormal basis in $\mathcal{H}_{\mathrm{kin}}^{\mathrm{geo}}$ by $\left\langle v_1 \mid v_2\right\rangle=\delta_{v_1, v_2}$. The polymerized momentum operator acts this basis as
\begin{align}
	\label{polymerized momentum operator1}
	\widehat{e^{i b}} \cdot | v \rangle=| v+2 \rangle.
\end{align}
On the other hand, the physical volume of the elemental cell is given by
\begin{align}
\label{physical volume}
V_{phy}=\int_{V} \sqrt{h} \ d^3 x=|p|^{2 / 3}=\frac{\sqrt{2 \sqrt{3}}}{8} \gamma^{3 / 2}|v|.
\end{align}
Since it is proportional to $|v|$, one can evaluate the evolution of the elemental cell by studying the evolution of $|v|$. 

For the scalar field part, we also employ the polymerlike quantization. The kinematical Hilbert space of the scalar field is constructed as $\mathcal{H}_{\mathrm{kin}}^{\mathrm{sc}}=L^2\left(R_{B o h r}, d \mu_H\right)$. The eigenstates of operator $\hat{\phi}$ corresponding to the scalar field are labeled by real numbers $\phi$ as
\begin{align}
\label{phi operator}
\hat{\phi} \cdot |\phi\rangle=\phi |\phi\rangle.
\end{align}
These states contribute an orthonormal basis in $\mathcal{H}_{\mathrm{kin}}^{\mathrm{sc}}$ by
\begin{align}
\label{basisphi}
\left\langle\phi_1 \mid \phi_2\right\rangle=\delta_{\phi_1, \phi_2},
\end{align}
where $\delta_{\phi_1, \phi_2}$ is the Kronecker delta function. The polymerized momentum operator acts this basis as 
\begin{align}
	\label{polymerized momentum operator2}
	\widehat{e^{i \lambda_0 \tilde{\pi}}} \cdot | \phi \rangle=| \phi-\lambda_0 \rangle.
\end{align}
The total kinematical Hilbert space of Brans-Dicke cosmology is constructed as the tensor product $\mathcal{H}_{\text {kin }}^{B D}=\mathcal{H}_{\mathrm {kin }}^{\mathrm{geo}} \otimes \mathcal{H}_{\mathrm {kin }}^{\mathrm{sc}}$.  

The deparametrized Hamiltonian constraint (\ref{new constraint1}) can be expressed by the basic variables as
\begin{align}
\label{rthe deparametrized Hamiltonian constraint}
C=\tilde{\pi}-\left(\frac{-3 \pm \sqrt{3(3+2 \omega)}}{4 \phi}\right) b v,
\end{align}
where the $+$ sign corresponds to the phase space region $\Gamma_{+}$ with $bv \geq 0$ and the $-$ sign corresponds to the region  $\Gamma_{-}$ with $bv \leq 0$. Classically, the choice of sign $\pm$ is determined by the region of phase space in which the initial state is selected. In the quantum theory, if one considers the evolution of a coherent state peaked at a point in phase space, the sign is set by the point at which the initial state is peaked. In what follows, for definiteness we will focus on the $+$ sign. Because of spatial homogeneity, the lapse function $N$ is taken to be one without loss of generality. Equation (\ref{rthe deparametrized Hamiltonian constraint}) can be regularized as
\begin{align}
\label{rregularizedconstraint}
C=&\frac{1}{2i \lambda_0}(e^{i \lambda_0 \tilde{\pi}}-e^{-i \lambda_0 \tilde{\pi}}) \nonumber \\ &-\left(\frac{-3 + \sqrt{3(3+2 \omega)}}{8 \phi}\right) \Big( (\sin b)  v+v  (\sin b) \Big),
\end{align}
where $\lambda_0$ is a small dimensionless constant. Then, it can be quantized as
\begin{align}
\label{rconstraintoperator}
\hat{C}=\frac{1}{2i \lambda_0}(\widehat{e^{i \lambda_0 \tilde{\pi}}}-\widehat{e^{-i \lambda_0 \tilde{\pi}}})-A\hat{\phi}^{-1} \hat{G},
\end{align}
where $A=\frac{-3 + \sqrt{3(3+2 \omega)}}{8}$ and $\hat{G}=(\widehat{\sin b})  \hat{v}+\hat{v}  (\widehat{\sin b})$. The operator $\hat{\phi}^{-1}$ is defined as
\begin{align}
\label{rphi-1}
\hat{\phi}^{-1} \cdot |\phi\rangle=\frac{4 \mathrm{sgn}(\phi)}{\lambda_0^2}\left(\sqrt{|\phi|}-\sqrt{\left|\phi+\lambda_0\right|}\right)^2|\phi\rangle \equiv D|\phi\rangle
\end{align}

\subsection{Solutions to the deparametrized constraint}
The physical states of the quantum theory have to satisfy the following quantum constraint equation:
\begin{align}
\label{quantum constraint equation}
\hat{C} \cdot \Psi(\phi, v)=0.
\end{align}
Note that one does not need to define the Hamiltonian constraint operator in the vertex Hilbert space, since the diffeomorphism constraint has been satisfied in the cosmological model. To solve Eq.~(\ref{quantum constraint equation}), we use the separation of variables by $\Psi(\phi, v)=U(\phi) \psi(v)$. Then, Eq.~(\ref{quantum constraint equation}) becomes
\begin{align}
\label{quantum constraint equation1}
\! \frac{1}{2 i \lambda_0} \! \left[U\left(\phi \!+\! \lambda_0\right) \!-\! U\left(\phi \!-\! \lambda_0\right)\right]\psi(v) \!-\! A \ D \ U(\phi)\hat{G} \psi(v) \!=\! 0,
\end{align}
which gives the following two equations:
\begin{align}
\label{quantum constraint equation2}
\frac{1}{2 i \lambda_0} \frac{1}{A \ D \ U(\phi)}\left[U\left(\phi+\lambda_0\right)-U\left(\phi-\lambda_0\right)\right]=k, \nonumber
\\ \frac{1}{\psi(v)} \hat{G} \psi(v)=k.
\end{align}
Since the left-hand side of the first equation is a function depending only on $\phi$ while the left-hand side of the second equation is a function depending only on $v$, $k$ is independent of both $\phi$ and $v$. Since the operator $\hat{G}$ is self-adjoint, $k$ takes values in $\mathbb{R}$. Since the effects of polymer quantization of the geometric part are primarily focused, we take $\lambda_0 \rightarrow 0$ in Eq.~(\ref{quantum constraint equation2}) to obtain an approximate differential equation for $U(\phi)$ as 
\begin{align}
\label{quantum constraint equation3}
\frac{1}{i A} \frac{\phi}{U(\phi)}\frac{dU(\phi)}{d\phi}=k.
\end{align}
Note that Eq.~(\ref{quantum constraint equation3}) can also be obtained by using the Schrödinger representation for the scalar field. Its solution reads
\begin{align}
\label{solutions1}
U(\phi)=e^{i A k \ln |\phi|}.
\end{align}
The second equation in Eq.~(\ref{quantum constraint equation2}) is the eigenequation for the operator $\hat{G}$, which can be written as
\begin{align}
\label{eigen equation ofG}
\hat{G} \cdot \psi(v)=i\left[(v+1) \psi(v+2)-(v-1) \psi(v-2)\right]=k \psi(v).
\end{align}
We denote its solution with eigenvalue $k$ as $\psi(v)=e_k(v)$. Then, the general solution to the quantum constraint (\ref{quantum constraint equation}) can be written as
\begin{align}
\label{general solutions}
\Psi(\phi, v)=\int_{-\infty}^{+\infty} d k \tilde{\Psi}(k) e_k(v) e^{i A k \ln |\phi|},
\end{align}
where $\tilde{\Psi}(k)$ belongs to $L^2(\mathbb{R}, d k)$. It can also be written in the evolutionary form as 
$\Psi(\phi, v)=e^{i A \hat{G}\left(\ln |\phi|-\ln |\phi_0|\right)} \Psi\left(\phi_0, v\right)$, where $\Psi\left(\phi_0, v\right)$ represents the initial data for the equation at time $\phi = \phi_0$. 

Notice that the "parity" operator can be defined as
\begin{align}
\label{parity}
\hat{\Pi} \cdot \psi(v)=\psi(-v).
\end{align}
Its commutation relation with the operator $\hat{G}$ can be calculated by
\begin{align}
\label{commutation relation}
& \hat{\Pi} \cdot \hat{G} \cdot \psi(v) \!=\! i[(-v+1) \psi(-v+2)\!-\!(-v-1) \psi(-v-2)], \nonumber\\
& \hat{G} \cdot \hat{\Pi} \cdot \psi(v) \!=\! i[(v+1) \psi(-(v+2))\!-\!(v-1) \psi(-(v-2))],
\end{align}
and, hence,
\begin{align}
\label{commutation relation1}
[\hat{G}, \hat{\Pi}]=0.
\end{align}
Therefore, one has
\begin{align}
\label{brelation}
\hat{G} \cdot e_k(-v)=\hat{\Pi} \cdot \hat{G} \cdot e_k(v)=k e_k(-v).
\end{align}
This implies that $e_k(-v)$ is also an eigenstate of operator $\hat{G}$ with eigenvalue $k$. Note that the operator $\hat{\Pi}$ is associated with reversing the triad orientation and has eigenvalues $\pm 1$. It would be useful for coupling fermions in the quantum theory. In this paper, we consider only the eigenspace of $\Pi$ with eigenvalue $+1$, which is the symmetric section\cite{ashtekar2006quantum}. 

Notice that Eq.~(\ref{eigen equation ofG}) takes the form of a recursion relation. Therefore, this equation can be solved by specifying two initial values $e_k(v_0)$ and $e_k(v_0 + 2)$ for a given $v_0$. The solution has support only at points $v_0 + 2n$ for $n \in \mathbb{Z}$. Hence, the space of physical states, i.e, appropriate solutions to the constraint equation, is naturally divided into infinite numbers of sectors, each of which is preserved by the "evolution" and by the action of Dirac observables. Thus, there is superselection\cite{ashtekar2006quantum}. Let $\mathcal{L}_{|\varepsilon|}$ denote the "lattice" of points $\{|\varepsilon|+2 n, n \in \mathbb{Z}\}$ with $-1 \leq \varepsilon \leq 1$ on the $v$ axis, and let $\mathcal{L}_{-|\varepsilon|}$ denote the lattice of points $\{-|\varepsilon|+2 n, n \in \mathbb{Z}\}$.  Define $\mathcal{L}_{\varepsilon}=\mathcal{L}_{|\varepsilon|} \bigcup \mathcal{L}_{-|\varepsilon|}$ so that $\mathbb{R}=\bigcup_{\varepsilon} \mathcal{L}_{\varepsilon}$. Let $\mathcal{H}^{BD}_{|\varepsilon|}$, $\mathcal{H}^{BD}_{-|\varepsilon|}$, and $\mathcal{H}^{BD}_{\varepsilon}$ denote the subspaces of $\mathcal{H}_{\mathrm{kin}}^{\mathrm{geo}}$ with states whose supports are restricted to $\mathcal{L}_{|\varepsilon|}$, $\mathcal{L}_{-|\varepsilon|}$, and $\mathcal{L}_{\varepsilon}$, respectively. On the one hand, since $\mathcal{H}^{BD}_{|\varepsilon|}$ and $\mathcal{H}^{BD}_{-|\varepsilon|}$ are mapped to each other by the parity operator $\hat{\Pi}$, only $\mathcal{H}^{BD}_{\varepsilon}$ is left invariant by $\hat{\Pi}$. On the other hand, we focus on the symmetric section. Therefore, we are particularly interested in the symmetric subspace of $\mathcal{H}^{BD}_{\varepsilon}$. If we have the eigenstates $e_k^{+}(v)$ and $e_k^{-}(v)$ in $\mathcal{H}^{BD}_{|\varepsilon|}$ and $\mathcal{H}^{BD}_{-|\varepsilon|}$, respectively, through Eq.~(\ref{eigen equation ofG}), the symmetric eigenstate in $\mathcal{H}^{BD}_{\varepsilon}$ can be given by 
\begin{align}
\label{symmetric eigenstate}
e_k^s(v):=e_k^{+}(v)+e_k^{+}(-v)+e_k^{-}(v)+e_k^{-}(-v),
\end{align}
where we have used Eq.~(\ref{brelation}). The choice of $\varepsilon$ can, in principle, be  determined experimentally,
provided one has access to microscopic measurements which can distinguish between values of the scale factor which differ by $\sim 1.3 \ell_{\mathrm{Pl}}$\cite{ashtekar2006quantum}. Of greater practical interest are the coarse-grained measurements, where the coarse graining occurs at significantly greater scales. For these measurements, different sectors would be indistinguishable, and one could work with any one\cite{ashtekar2006quantum}. For simplicity, we choose $\varepsilon=0$ for our calculations. A symmetric eigenstate of $\hat{G}$ is shown in Fig. 1, where only the sector with $v \geq 0$ is displayed due to the symmetry.
\begin{figure}
\centering
\includegraphics[height=4.5cm,width=8.5cm]{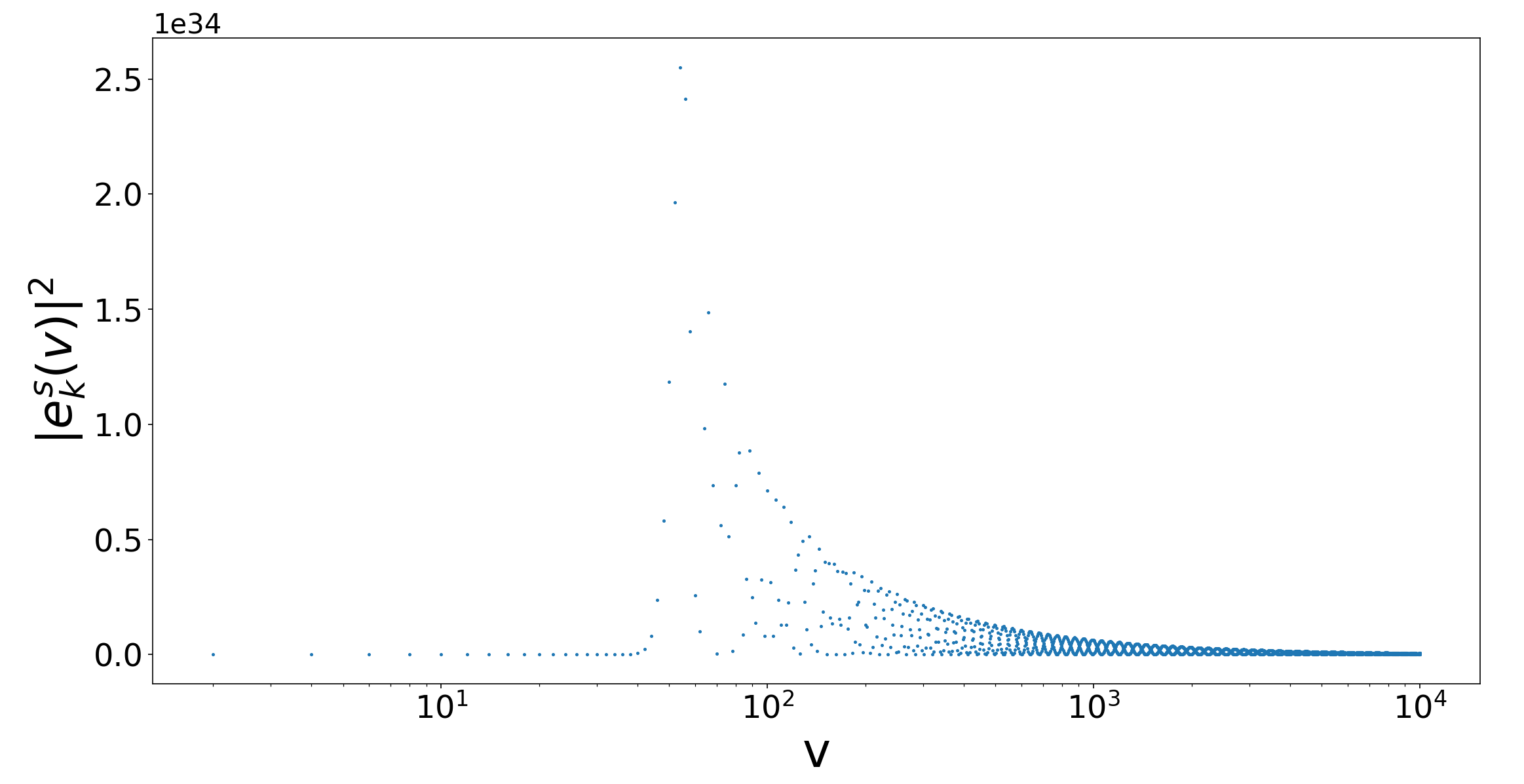}
\caption{The plot of $|e_k^{s}(v)|^2$ for an symmetric eigenstate $e_k^{s}(v)$ of operator $\hat{G}$ with the parameter is chosen as $k=100$: it shows that this eigenstate is not normalized.}
\end{figure}

\subsection{Normalization of the symmetric eigenstates}
It is necessary to do normalization for the symmetric eigenstate of the operator $\hat{G}$. However, because the symmetric eigenstate does not decay quickly enough, it cannot be normalized only through numerical calculations. Thus, we need to find its asymptotic analytic approximation. The strategy is to derive an approximate differential equation of Eq.~(\ref{eigen equation ofG}) as $v \rightarrow \infty$. However, an approximate differential equation cannot be obtained directly, because the symmetric eigenstate is not slowly varying. As shown in Fig. 2, there is a significant difference between the values of any two adjacent points of the symmetric eigenstate.
\begin{figure}
\centering
\includegraphics[height=4.5cm,width=8.5cm]{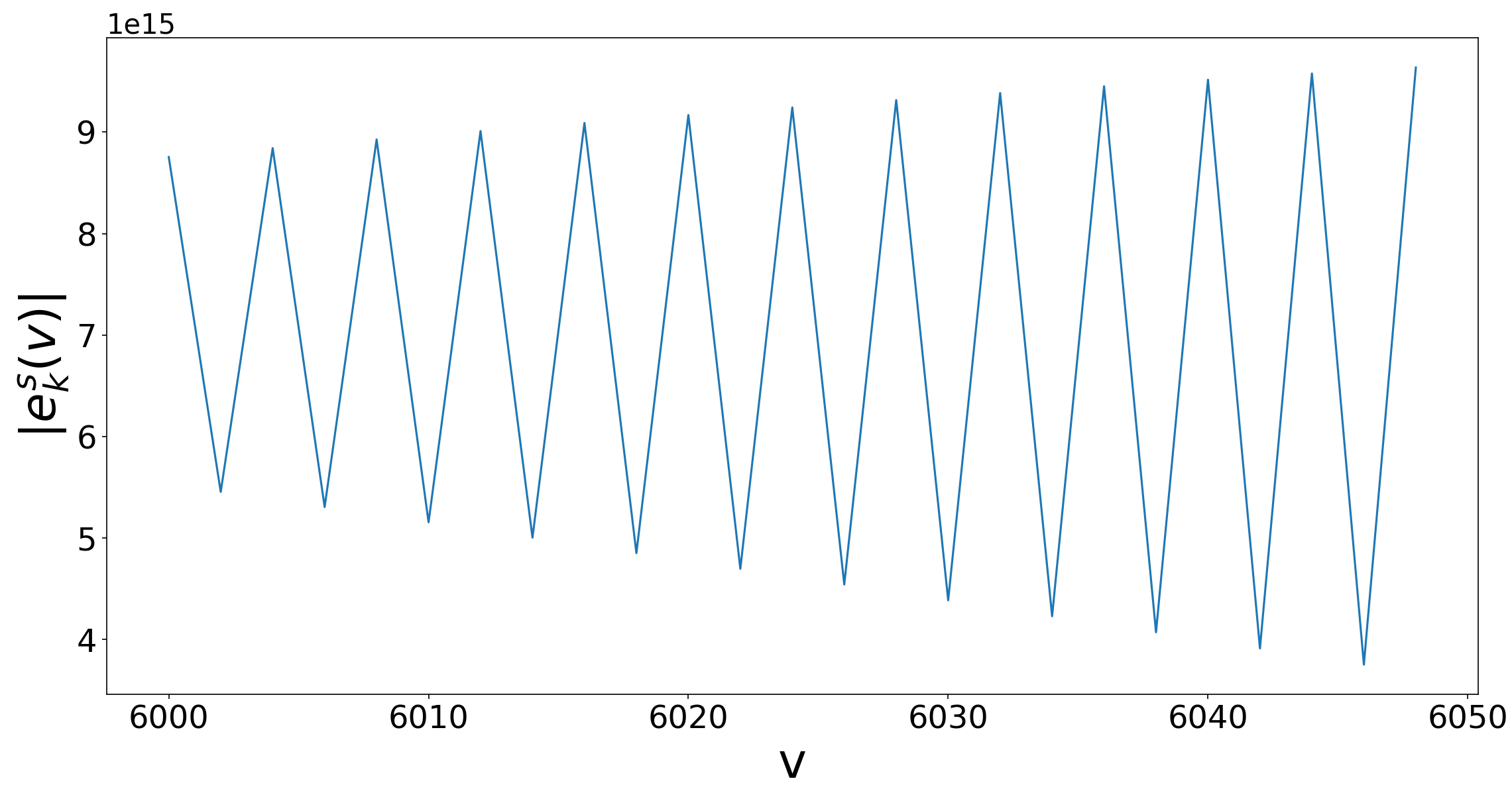}
\caption{The plot of the local graph of $|e_k^{s}(v)|$ for an symmetric eigenstate $e_k^{s}(v)$ of operator $\hat{G}$ with the parameter $k=100$: the values of $e_k^{s}(v)$ vary significantly between two adjacent lattice points, and, hence, it is not possible to approximate this eigenstate with a smooth analytical function.}
\end{figure}
Nevertheless, the difference between the values of the two points separated by an intermediate point is small. This leads us to divide the symmetric eigenstate into two sectors
\begin{equation}
	\begin{aligned}
		\label{eigenstate into two sectors}
		e_k^{s}(v)=\left\{\begin{array}{l}
			\hat{e}_k^{s}(v), \ \ \ \ \ \ \text{when} \ v=4 n \pm |\varepsilon|, \\
			\tilde{e}_k^{s}(v), \ \ \ \ \ \ \text{when} \ v=4 n+2 \pm |\varepsilon|,
		\end{array}\right.
	\end{aligned}
\end{equation}
where $n \in \mathbb{Z}$. As shown in Fig. 3, the function $e_k^{s}(v)$ varies slowly within each sector.
\begin{figure}
\centering
\includegraphics[height=4.5cm,width=8.5cm]{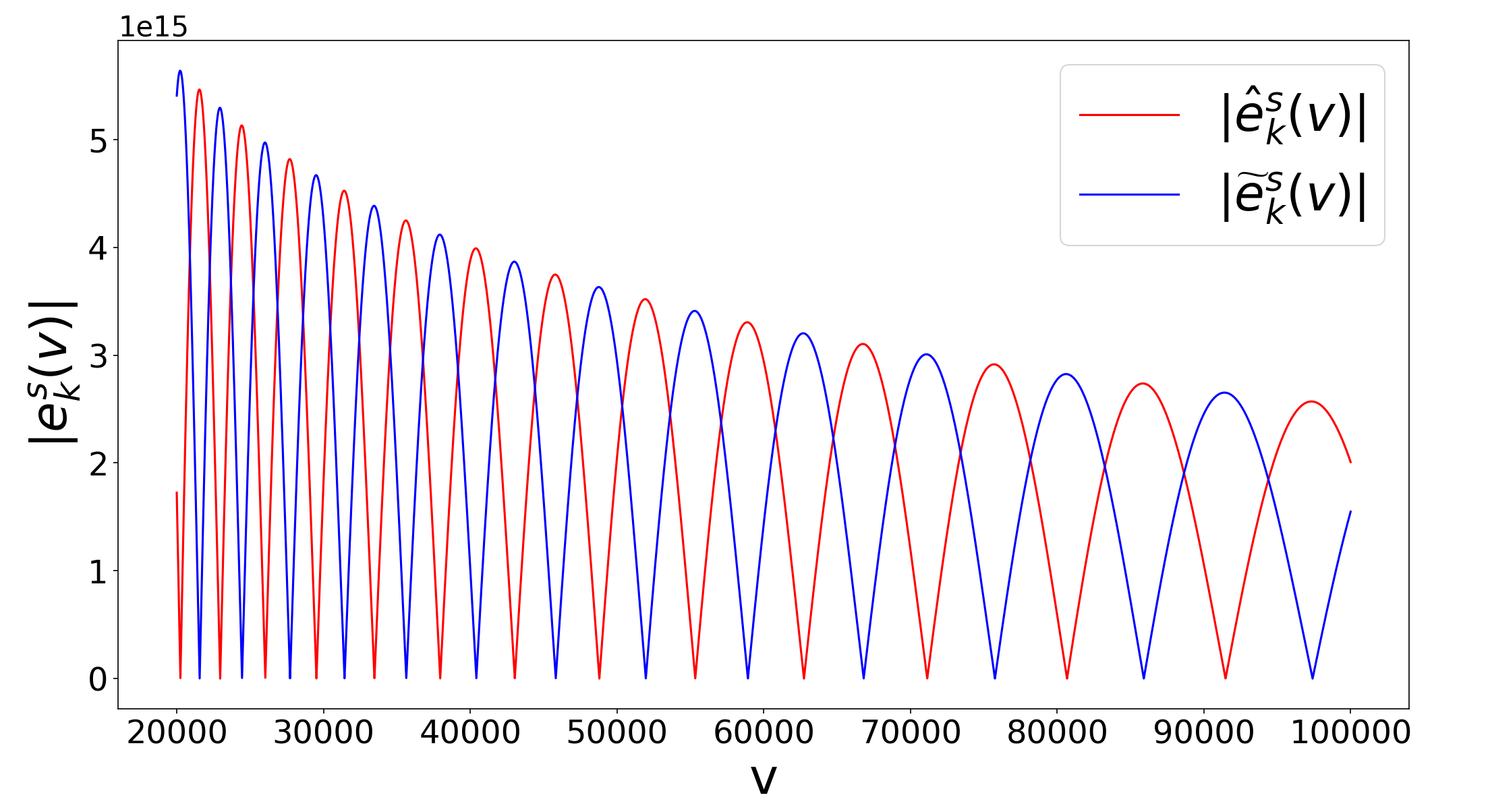}
\caption{The plot of $|e_k^{s}(v)|$ for an symmetric eigenstate $e_k^{s}(v)$ of operator $\hat{G}$ with into two sectors and with the parameter $k=100$: it shows that the eigenstate $e_k^{s}(v)$ tends to two distinct differentiable functions at large $v$.}
\end{figure}

Since the function $e_k^{s}(v)$ satisfies the eigenequation (\ref{eigen equation ofG}), we have
\begin{align}
\label{twoeigen equation}
(v+1)\tilde{e}_k^{s}(v+2)- (v-1)\tilde{e}_k^{s}(v-2)=-i k \hat{e}_k^{s}(v), \nonumber
\\ (v-1)\hat{e}_k^{s}(v)- (v-3)\hat{e}_k^{s}(v-4)=-i k \tilde{e}_k^{s}(v-2),
\end{align}
where $v=4n \pm |\varepsilon|$. Let $\hat{\psi}(v)$ and $\tilde{\psi}(v)$ be the approximate smooth functions of $\hat{e}_k^{s}(v)$ and $\tilde{e}_k^{s}(v)$, respectively. By using Eq.~(\ref{twoeigen equation}), they satisfy
\begin{align}
\label{tapproximate equation}
4(v+1)\tilde{\psi}^{\prime}(v-2)+2\tilde{\psi}(v-2)=-i k \hat{\psi}(v), \nonumber
\\ 4(v-1)\hat{\psi}^{\prime}(v-4)+2\hat{\psi}(v-4)=-i k \tilde{\psi}(v-2),
\end{align}
where a prime detotes the derivative with respect to $v$. Taking the derivative with respect to $v$ for the second equation in Eq.~(\ref{tapproximate equation}) and then combining it with Eq.~(\ref{tapproximate equation}), we obtain
\begin{align}
\label{fapproximate equation}
&(16v^2+128v+240)\hat{\psi}^{\prime \prime}(v)  +(32v+4k^2+144)\hat{\psi}^{\prime}(v)\nonumber \\&+(k^2+4)\hat{\psi}(v)=0.
\end{align}
Its asymptotically approximate differential equation as $v \rightarrow \infty$ reads
\begin{align}
\label{aapproximate equation}
(16v^2)\hat{\psi}^{\prime \prime}(v)+(32v)\hat{\psi}^{\prime}(v)+(k^2+4)\hat{\psi}(v)=0.
\end{align}
The solution to Eq.~(\ref{aapproximate equation}) is of the form  
\begin{align}
\label{solution of aapproximate equation}
\hat{\psi}(v)=C_1 \frac{1}{\sqrt{|v|}} e^{i \frac{1}{4} |k| \ln |v|}+C_2 \frac{1}{\sqrt{|v|}} e^{-i \frac{1}{4} |k| \ln |v|},
\end{align}
where the constants $C_1$ and $C_2$ are determined by the initial conditions. The same method can be applied to $\tilde{\psi}(v)$ to achieve the same result. Denoting $\underline{e}_{|k|}(v) \equiv \frac{1}{\sqrt{2\pi}} \frac{1}{\sqrt{|v|}} e^{i \frac{1}{4} |k| \ln |v|}$ and $\underline{e}_{-|k|}(v) \equiv \frac{1}{\sqrt{2\pi}} \frac{1}{\sqrt{|v|}} e^{-i \frac{1}{4} |k| \ln |v|}$, the symmetric eigenstate can be written as
\begin{align}
\label{asymptotic eigenstate1}
\lim _{v \rightarrow \infty} \hat{e}_k^s(v)=r^{+} e^{i \alpha} \underline{e}_{|k|}(v)+r^{-} e^{i \beta} \underline{e}_{-|k|}(v) \equiv \hat{\psi}_k(v),
\end{align}
\begin{align}
\label{asymptotic eigenstate2}
\lim _{v \rightarrow \infty} \tilde{e}_k^s(v)=\tilde{r}^{+} e^{i \tilde{\alpha}} \underline{e}_{|k|}(v)+\tilde{r}^{-} e^{i \tilde{\beta}} \underline{e}_{-|k|}(v) \equiv \tilde{\psi}_k(v).
\end{align}
Multiplying by $\sqrt{|v|}$ in Eqs.~(\ref{asymptotic eigenstate1}) and (\ref{asymptotic eigenstate2}) and applying the method introduced in Appendix B in Ref.\cite{ashtekar2006quantum}, the values of the parameters $r^{+}, r^{-}, \alpha, \beta$, $\tilde{r}^{+}, \tilde{r}^{-}, \tilde{\alpha}$ and $\tilde{\beta}$ can be determined. A symmetric eigenstate is compared with its asymptotic approximation state in Figs. 4--7.
\begin{figure}
\centering
\includegraphics[height=4.5cm,width=8.5cm]{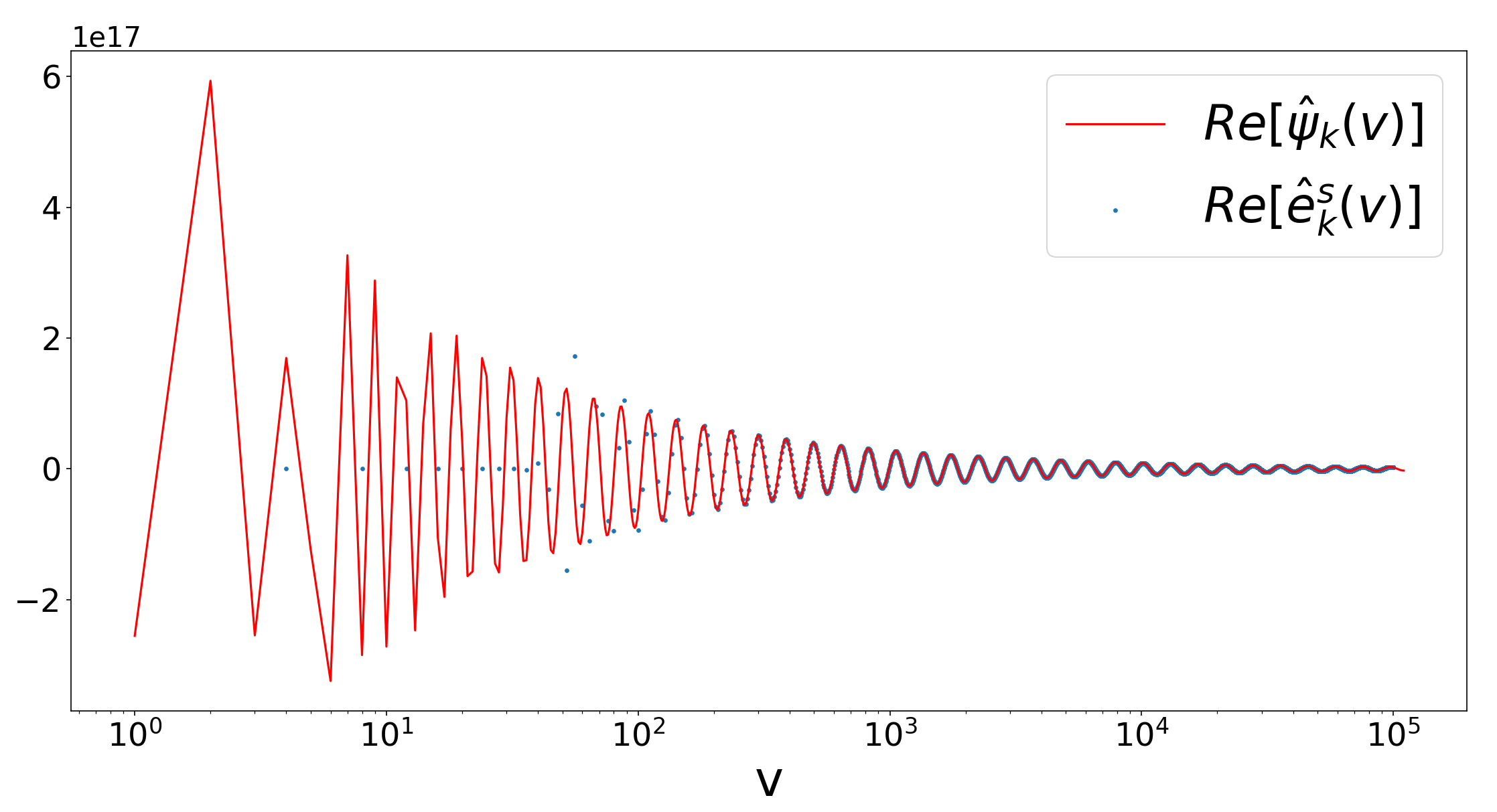}
\caption{The real part of the symmetric eigenstate $e_k^{s}(v)$ in the sector supported on $v=4n$ (denoted by the blue point) compared with the real part of its asymptotically approximate function (denoted by the red solid line) with the parameter $k=100$: this demonstrates a very good agreement between the two functions at large $v$.}
\end{figure}
\begin{figure}
\centering
\includegraphics[height=4.5cm,width=8.5cm]{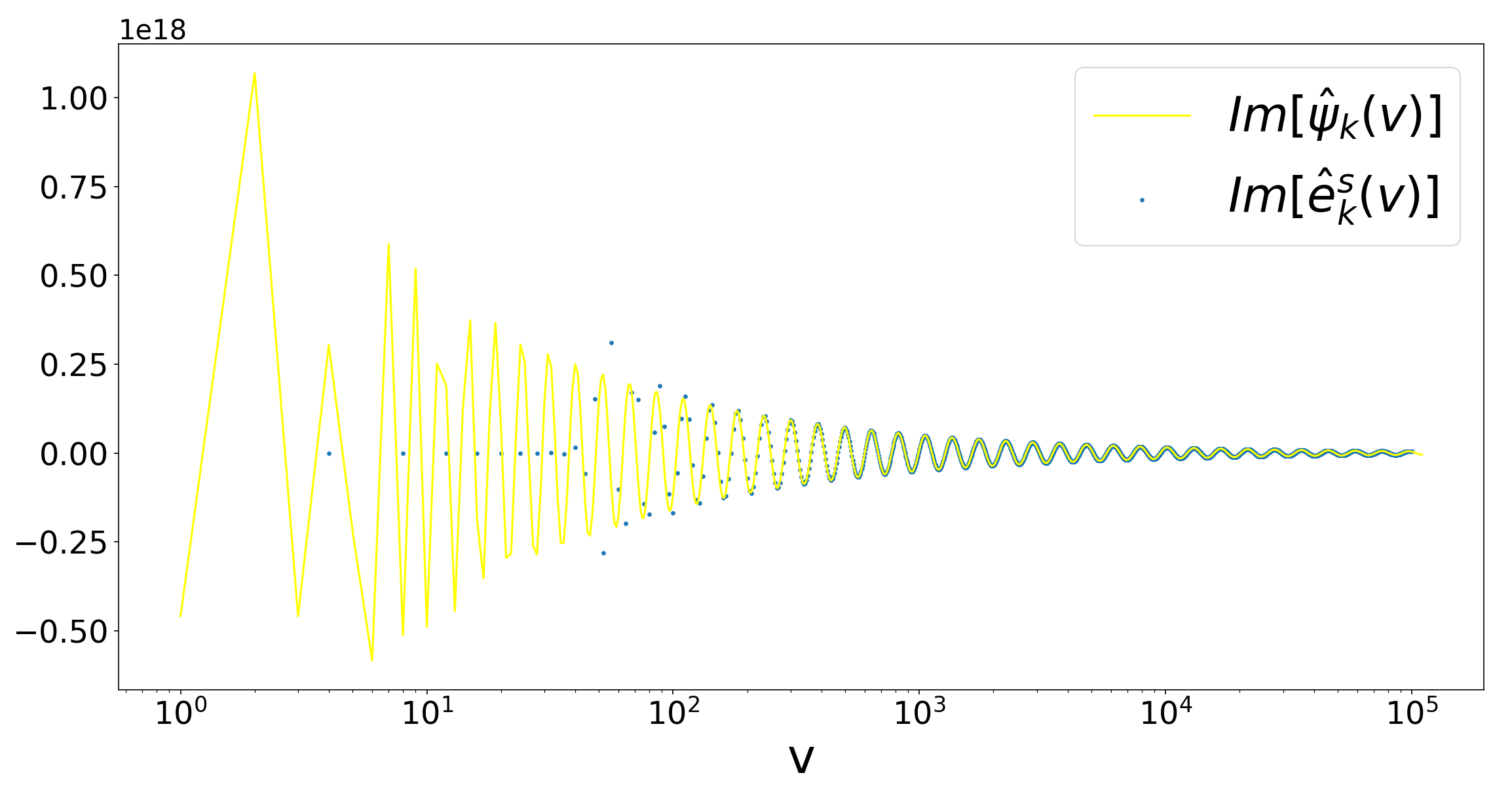}
\caption{The image part of the symmetric eigenstate $e_k^{s}(v)$ in the sector supported on $v=4n$ (denoted by the blue point) compared with the image part of its asymptotically approximate function (denoted by the yellow solid line) with the parameter $k=100$: this demonstrates a very good agreement between the two functions at large $v$.}
\end{figure}
\begin{figure}
\centering
\includegraphics[height=4.5cm,width=8.5cm]{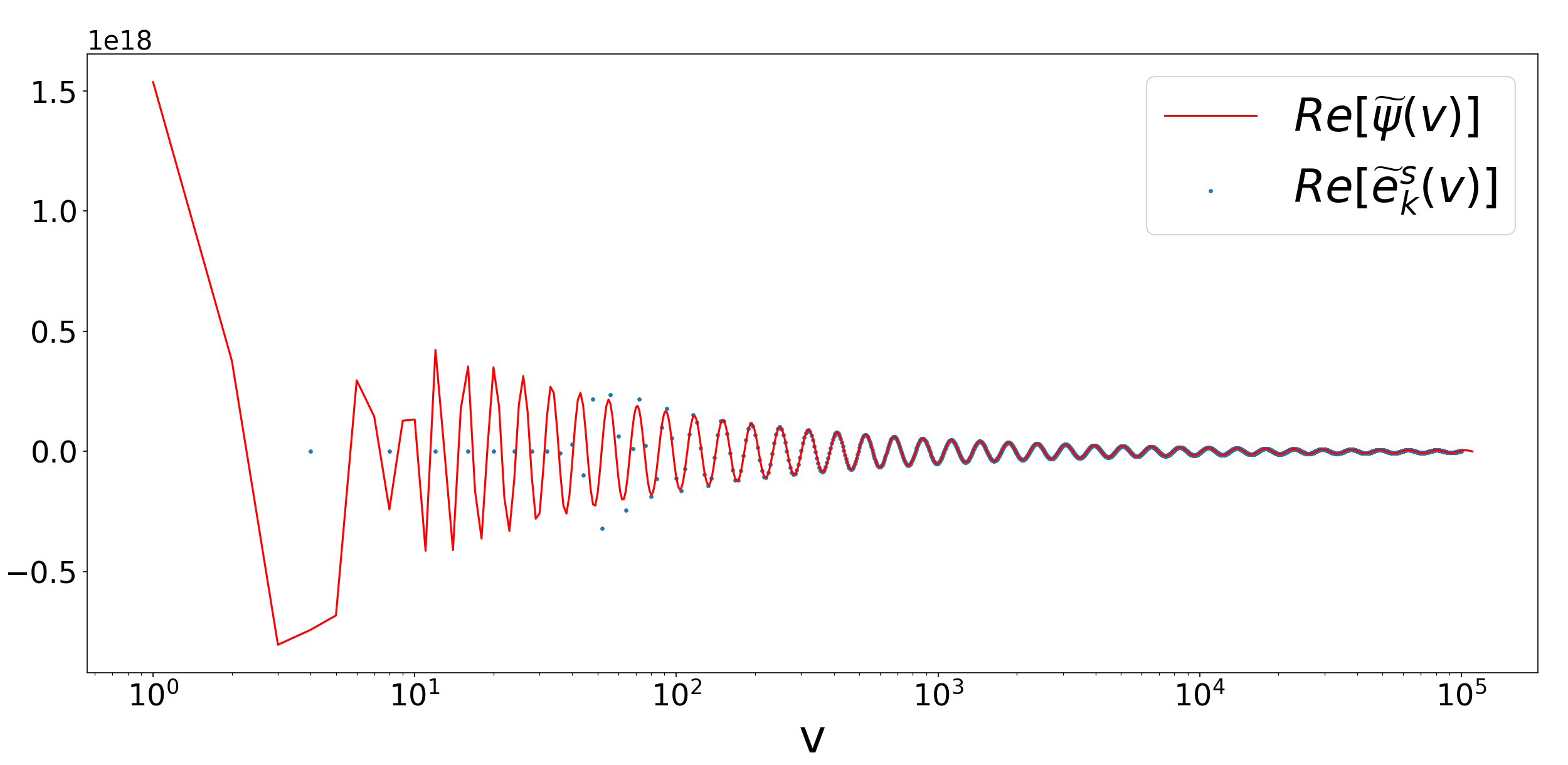}
\caption{The real part of the symmetric eigenstate $e_k^{s}(v)$ in the sector supported on $v=4n+2$ (denoted by the blue point) compared with the real part of its asymptotically approximate function (denoted by the red solid line) with the parameter $k=100$: this demonstrates a very good agreement between the two functions at large $v$.}
\end{figure}
\begin{figure}
\centering
\includegraphics[height=4.5cm,width=8.5cm]{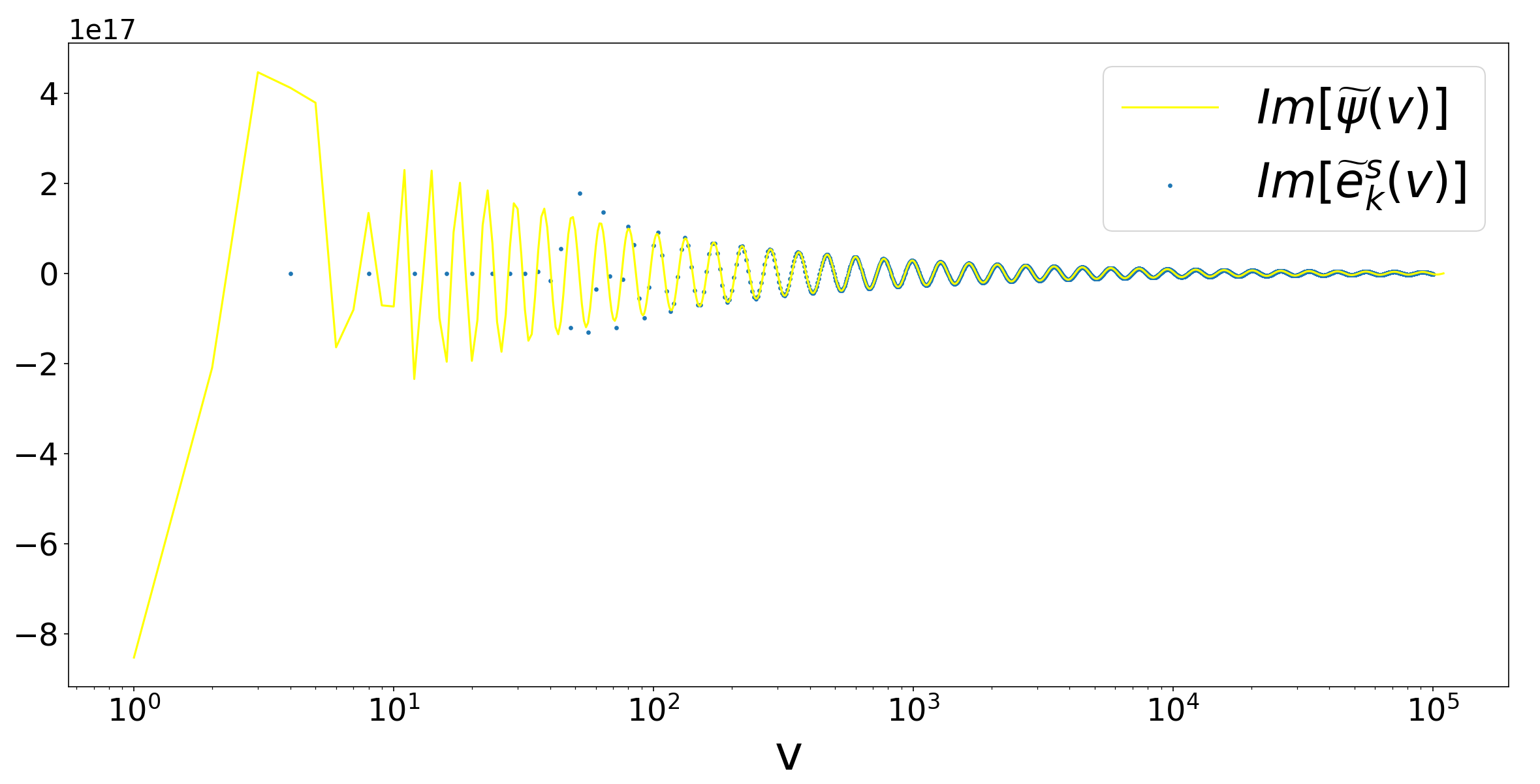}
\caption{The image part of the symmetric eigenstate $e_k^{s}(v)$ in the sector supported on $v=4n+2$ (denoted by the blue point) compared with the image part of its asymptotically approximate function (denoted by the yellow solid line) with the parameter $k=100$: this demonstrates a very good agreement between the two functions at large $v$.}
\end{figure}
It is easy to see that the function $\underline{e}_{|k|}(v)$ is not square integrable, and, hence, the state $e_{k}^{s}(v)$ cannot be normalized. What we can do is to normalize the symmetric eigenstates by the Dirac delta function. By the inner product on $\mathcal{H}^{BD}_{\varepsilon}$ induced from $\mathcal{H}_{\mathrm{kin}}^{\mathrm{geo}}$, one has
\begin{align}
\label{normalize}
X\left(k, k^{\prime}\right):=&\left\langle e_k^s \mid e_{k^{\prime}}^s\right\rangle=\sum_{L_{\varepsilon}} \overline{e_k^s(v)} e_{k^{\prime}}^s(v) \nonumber \\=&\sum_{\hat{L}_{\varepsilon}} \overline{\hat{e}_k^s(v)} \hat{e}_{k^{\prime}}^s(v)+\sum_{\tilde{L}_{\varepsilon}} \overline{\tilde{e}_k^s(v)} \tilde{e}_{k^{\prime}}^s(v),
\end{align}
where the sum is divided into two sectors $\hat{L}_{\varepsilon}=L_{\varepsilon} \cap \{4n \pm |\varepsilon| \}$ and $\tilde{L}_{\varepsilon}=L_{\varepsilon} \cap \{4n+2 \pm |\varepsilon| \}$. The first term $\hat{X}\left(k, k^{\prime}\right) \equiv \sum_{\hat{L}_{\varepsilon}} \overline{\hat{e}_k^s(v)} \hat{e}_{k^{\prime}}^s(v)$ in the right-hand side of Eq.(\ref{normalize}) can be written as
\begin{align}
\label{normalize2}
\hat{X}\left(k, k^{\prime}\right)&=2F_1(k,k^\prime)+2 \sum_{L^1} \overline{\hat{e}_k^s(v)} \hat{e}_{k^{\prime}}^s(v) \nonumber
\\ &\!=\! F_2\left(\! k, k^{\prime}\! \right) \!+\! 2 \! \sum_{L^1} r^{+} \! r^{+ \prime} \! e^{i\left(\alpha^{\prime}\! -\! \alpha\right)} \underline{e}_{-|k|}(v) \! \underline{e}_{\left|k^{\prime}\right|}(v) \nonumber\\
& +2 \sum_{L^1} r^{+} r^{- \prime} e^{i\left(\beta^{\prime}-\alpha\right)} \underline{e}_{-|k|}(v) \underline{e}_{-\left|k^{\prime}\right|}(v) \nonumber\\
&  +2 \sum_{L^1} r^{-} r^{+\prime} e^{i\left(\alpha^{\prime}-\beta\right)} \underline{e}_{|k|}(v) \underline{e}_{\mid k^{\prime} \mid}(v) \nonumber\\
&  +2 \sum_{L^1} r^{-} r^{- \prime} e^{i\left(\beta^{\prime}-\beta\right)} \underline{e}_{\mid k \mid}(v) \underline{e}_{-\left|k^{\prime}\right|}(v) \nonumber 
 \\&\!=\! F_3\left(k, k^{\prime}\right) \!+\! \frac{1}{2} r^{+} \! r^{+\prime} \! e^{i\left(\alpha^{\prime} \!-\! \alpha\right)} \int_1^{\infty} \! \underline{e}_{-|k|}(v) \! \underline{e}_{\left|k^{\prime}\right|}(v) \nonumber\\
&  +\frac{1}{2} r^{+} r^{- \prime} e^{i\left(\beta^{\prime}-\alpha\right)} \int_1^{\infty} \underline{e}_{-|k|}(v) \underline{e}_{-\left|k^{\prime}\right|}(v) \nonumber \\
&  +\frac{1}{2} r^{-} r^{+ \prime} e^{i\left(\alpha^{\prime}-\beta\right)} \int_1^{\infty} \underline{e}_{|k|}(v) \underline{e}_{\left|k^{\prime}\right|}(v) \nonumber \\
&  +\frac{1}{2} r^{-} r^{- \prime } e^{i\left(\beta^{\prime}-\beta\right)} \int_1^{\infty} \underline{e}_{|k|}(v) \underline{e}_{-\left|k^{\prime}\right|}(v) \nonumber
\\& = F_4\left(k, k^{\prime}\right) +\frac{1}{4} r^{+} r^{+\prime} e^{i\left(\alpha^{\prime}-\alpha\right)} 4 \delta\left(\left|k^{\prime}\right|-|k|\right) \nonumber \\
& +\frac{1}{4} r^{+} r^{-\prime} e^{i\left(\beta^{\prime}-\alpha\right)} 4 \delta\left(\left|k^{\prime}\right|+|k|\right) \nonumber \\
& +\frac{1}{4} r^{-} r^{+\prime} e^{i\left(\alpha^{\prime}-\beta\right)} 4 \delta\left(\left|k^{\prime}\right|+|k|\right) \nonumber \\
& +\frac{1}{4} r^{-} r^{-\prime} e^{i\left(\beta^{\prime}-\beta\right)} 4 \delta\left(\left|k^{\prime}\right|-|k|\right),
\end{align}
where we denote $L^1 \equiv \left.\hat{L}_{\varepsilon}\right|_{v \geqslant 1}$ and $F_1(k,k^{\prime})$, $F_2(k,k^{\prime})$, $F_3(k,k^{\prime})$, and $F_4(k,k^{\prime})$ are well-defined functions with finite values. Note that the sum over region of $v \geqslant 1$ is split in the first step of Eq.(\ref{normalize2}), Eq.~(\ref{asymptotic eigenstate1}) is used in the second step, the sums are estimated via integrals in the third step, and the results of the integrals are applied in the last step. A more detailed discussion on this normalization method can be found in Appendix A in Ref.\cite{kaminski2010cosmic}. Equation (\ref{normalize2}) implies
\begin{align}
\label{normalize3}
\hat{X}\left(k, k^{\prime}\right) \! \approx \! \left(r^{+} r^{+\prime} e^{i\left(\alpha^{\prime}-\alpha\right)}+r^{-} r^{- \prime} e^{i\left(\beta^{\prime}-\beta\right)}\right) \delta\left(\left|k^{\prime}\right|-|k|\right),
\end{align}
where $\approx$ means being equal up to a well-defined function. We denote the second term in the right-hand side of Eq.(\ref{normalize}) by $\tilde{X}\left(k, k^{\prime}\right) \equiv \sum_{\hat{L}_{\varepsilon}} \overline{\tilde{e}_k^s(v)} \tilde{e}_{k^{\prime}}^s(v)$. Similarly, it can be estimated as
\begin{align}
\label{normalize4}
\tilde{X}\left(k, k^{\prime}\right) \! \approx \! \left(\tilde{r}^{+} \tilde{r}^{+\prime} e^{i\left(\tilde{\alpha}^{\prime}-\tilde{\alpha}\right)}+\tilde{r}^{-} \tilde{r}^{- \prime} e^{i\left(\tilde{\beta}^{\prime}-\tilde{\beta}\right)}\right) \delta\left(\left|k^{\prime}\right|-|k|\right).
\end{align}
Therefore, the normalized symmetry eigenstate can be defined as
\begin{equation}
\begin{aligned}
\label{normalize5}
\breve{e}_k^s(v)=e_k^s(v) / \sqrt{(r^+)^2+(r^-)^2+(\tilde{r}^+)^2 +(\tilde{r}^-)^2}.
\end{aligned}
\end{equation}
A normalized symmetry eigenstate is shown in Fig. 8.
\begin{figure}
\centering
\includegraphics[height=4.5cm,width=8.5cm]{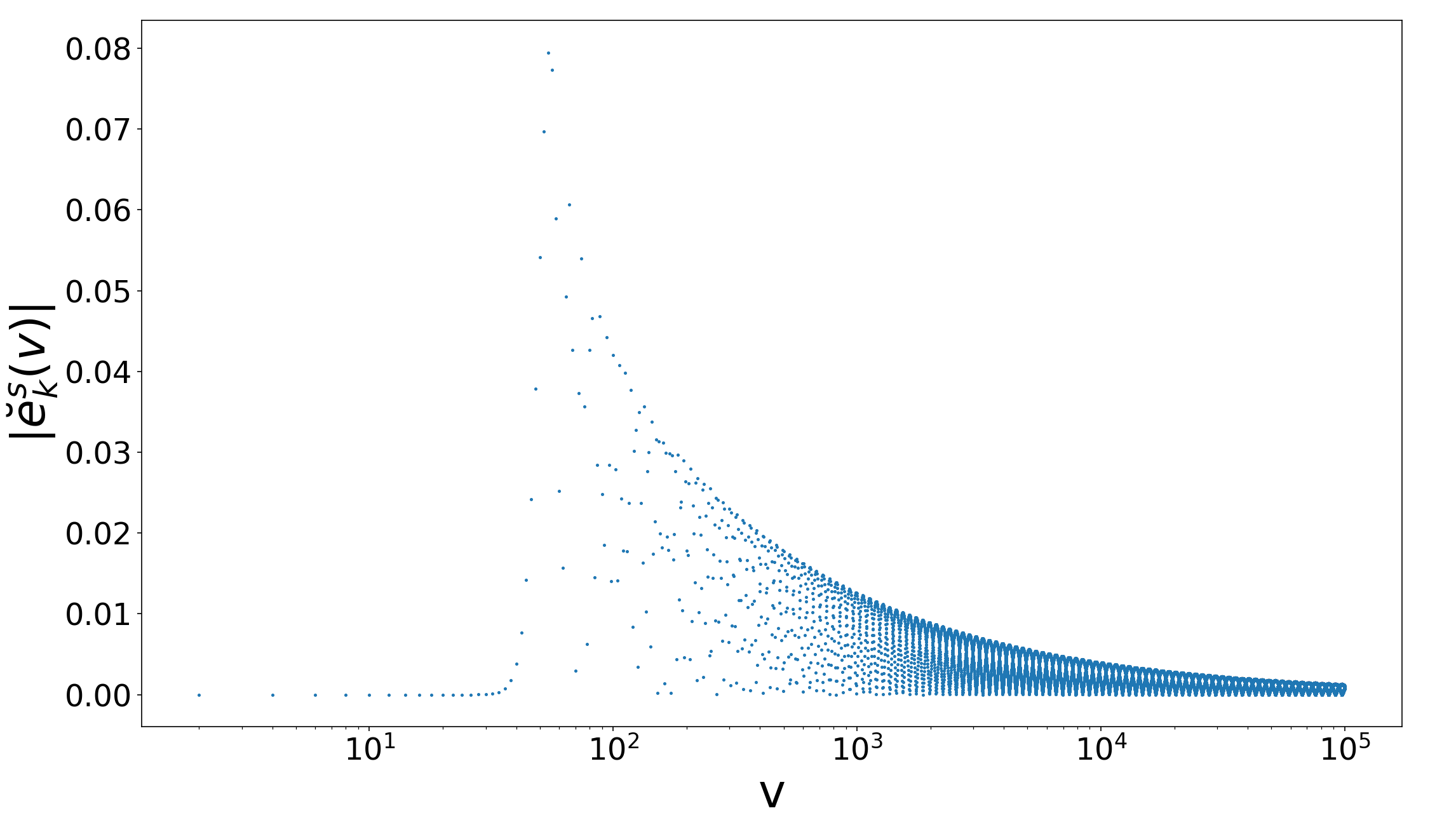}
\caption{The plot of $|\breve{e}_k^{s}(v)|$ for a normalized symmetric eigenstate $\breve{e}_k^s(v)$ of operator $\hat{G}$ with the parameter $k=100$: it differs from the unnormalized symmetric eigenstate $e_k^{s}(v)$ in Fig.1 by just a normalization factor.}
\end{figure}
\subsection{The physical sector, coherent state, and quantum bounce}
The physical Hilbert space $\mathcal{H}_{\varepsilon}^{phy}$ consists of the kernel of the Hamiltonian constraint operator $\hat{C}$ with initial data in the symmetric sector of $\mathcal{H}_{\varepsilon}^{B D}$. Thus, the elements in $\mathcal{H}_{\varepsilon}^{phy}$ can be expressed as
\begin{align}
\label{physical Hilbert space}
\Psi(\phi, v)=\int_{-\infty}^{+\infty} d k \tilde{\Psi}(k) \breve{e}^{s}_{k}(v) e^{i A k \ln |\phi|}.
\end{align}
The inner product on $\mathcal{H}_{\varepsilon}^{phy}$ is induced from that on $\mathcal{H}_{\varepsilon}^{B D}$ by 
\begin{align}
\label{bdphysical inner product}
\left\langle\Psi_1 \mid \Psi_2\right\rangle_{\varepsilon}=\sum_{v \in\{ \pm|\varepsilon|+2 n ; n \in \mathbb{Z}\}} \bar{\Psi}_1\left(v, \phi_o\right) \Psi_2\left(v, \phi_o\right)
\end{align}
for any $\phi_0$. In order to study the evolution of the volume of the elemental cell, we consider the quantum Dirac observable corresponding to $|v|$, which is defined by $\widehat{|v|_{\phi_0}}:=e^{i A \hat{G}\left(\ln |\phi|-\ln |\phi_0|\right)}|\hat{v}| e^{-i A \hat{G}\left(\ln |\phi|-\ln |\phi_0|\right)}$. A straightforward calculation shows it commute with the Hamiltonian constraint
\begin{align}
	\label{voperator0}
	[\widehat{|v|}_{\phi_0},\hat{C}]=0.
\end{align}
This shows that $\widehat{|v|_{\phi_0}}$ is a quantum Dirac observable. Note that it acts on the physical states as
\begin{align}
\label{voperator}
\widehat{|v|}_{\phi_0} \Psi(\phi, v)=e^{i A \hat{G}\left(\ln |\phi|-\ln |\phi_0|\right)}|v| \Psi\left(\phi_0, v\right).
\end{align}
Its expectation value under a physical state can be expressed as
\begin{align}
\label{vexpectation values}
\left\langle\Psi \mid \widehat{|v|}_{\phi_0} \mid \Psi\right\rangle_{\varepsilon}=\sum_{v \in\{ \pm|\varepsilon|+2 n ; n \in \mathbb{Z}\}} \bar{\Psi}\left(v, \phi_o\right) |v| \Psi \left(v, \phi_o\right).
\end{align}
To understand the evolutional behavior of semiclassical states, we construct a physical  coherent state as
\begin{align}
\label{coherent state}
\Psi(\phi, v)=\int_{-\infty}^{+\infty} d k e^{-\frac{(k-k^{\star})^2}{2 \sigma^2}} \breve{e}^{s}_{k}(v) e^{i A k (\ln         |\phi|- \ln |\phi^{\star}|)},
\end{align}
where $\sigma$ is the Gaussian spread. Note that this coherent state is peaked at $G=k^{\star}$, while at certain time $\phi=\phi_0$ it is also picked at $|v|_{\phi_ 0}=v^{\star}$ which is determined by $\phi^{\star}$. The expectation values $\langle \widehat{|v|}_{\phi_0} \rangle$ of operator $\widehat{|v|}_{\phi_0}$ under the coherent state (\ref{coherent state}) can be calculated by the method of fast Fourier transform and its inverse\cite{ashtekar2006quantum}. As shown in Fig. 9, there is a quantum bounce resolving the classical big bang singularity. 
\begin{figure}
\centering
\includegraphics[height=4.5cm,width=8.5cm]{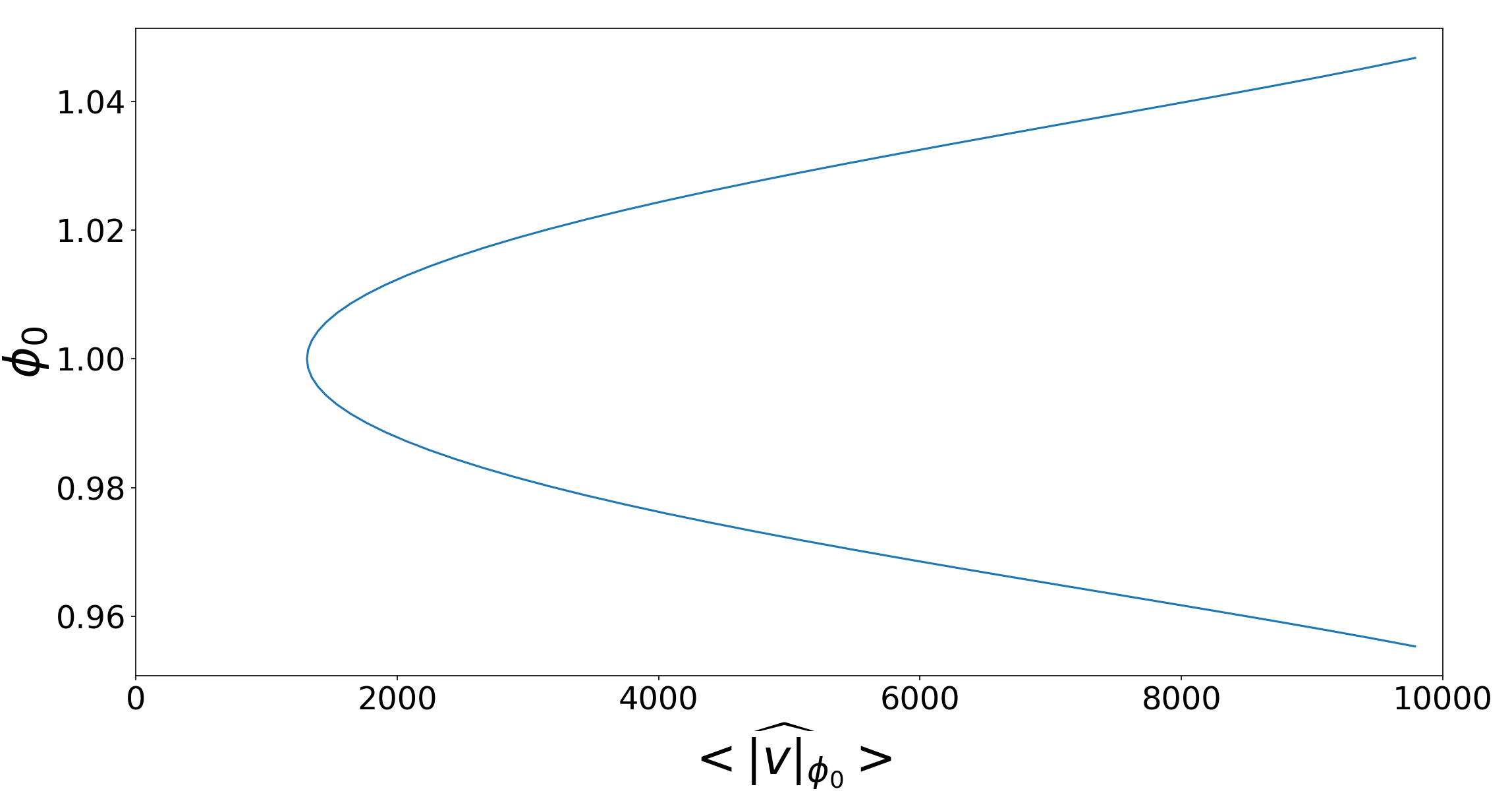}
\caption{The expectation values of operator $\widehat{|v|}_{\phi_0}$ under the coherent state (\ref{coherent state}) with the parameters $\omega=10000, \sigma=1, k^{\star}=100$, and $\phi^{\star}=1$: the classical big bang singularity is resolved by a quantum bounce.}
\end{figure}

\section{Conclusion}
\label{so}

In summarize, the Hamiltonian constraint of scalar-tensor theory of gravity has been deparamatrized by the degree of freedom of the scalar field and then loop quantized in Sec. III. In particular, in the sector of $\omega(\phi) = -\frac{3}{2}$, the deparametrization was carried out by combining the conformal constraint and the Hamiltonian constraint. In both sectors of $\omega(\phi) \neq -\frac{3}{2}$ and $\omega(\phi) = -\frac{3}{2}$, the deparametrized Hamiltonian constraints were promoted to well-defined operators on the vertex Hilbert space $\mathcal{H}_{vtx}$. The deparametrization and loop quantization approach of the full theory has also been applied to the model of spatially flat, homogeneous, and isotropic Brans-Dicke cosmology of the sector $\omega(\phi) \neq -\frac{3}{2}$ in Sec. IV. The Hamiltonian constraint in this cosmological model was deparametrized by the degree of freedom of the scalar and then loop quantized. The general solution to the quantum constraint equation was obtained by using the method of separation of variables, since the constraint could be divided into two difference equations, one of which is the eigenequation of the self-adjoint operator $\hat{G}$. It turns out that the symmetric eigenstates of $\hat{G}$ can be normalized by the Dirac delta function through the asymptotic analytic approximation. The quantum dynamics of the cosmological model was revealed by calculating numerically the expectation value of the operator $\widehat{|v|}_{\phi_0}$ under the physical coherent states. It turns out that the classical big bang singularity is replaced by a quantum bounce. 

The above results show that the degree of freedom of the scalar field can play a key role in the deparametrization of the Hamiltonian constraint in scalar-tensor gravity at both classical and quantum levels. It gives a discrete time evolution of the physical states with respect to the gravitational degree of freedom. Before, this kind of evolution was realized only in the $U(1)^3$ model\cite{Bakhoda2025u13}. In the cosmological model, physical solutions to the constraint have been successfully obtained in the light of the deparametrization. Nevertheless, there are still a few open issues that deserve further investigation. For example, the general solution to the deparametrized Hamiltonian constraint in the full theory is unknown, the conformal constraint has not been solved, and the value of the regularization parameter $\lambda_0$ is yet to be determined, and so on.

\section{ACKNOWLEDGMENTS}
\label{a}
We thank Cong Zhang, Shupeng Song  and Shun Jiang for helpful discussions. This work is supported by the National Natural Science Foundation of China (Grant No.12275022). 

\section{DATA AVAILABILITY}
The data that support the findings of this article are openly available in Ref.\cite{Yuan2026data}.
%%%%%%%%%%%%%%%%%%%%%%%%%%%%%%%%%%%%%%%%%%%%%%
%%%%%%%%%%%%%%%%%%%%%%%%%%%%%%%%%%%%%%%%%%%%%%
%%%%%%%%%%%%%%%%%%%%%%%%%%%%%%%%%%%%%
\bibliographystyle{amsalpha}

\end{document}